\documentclass[twocolumn,trackchanges]{aastex63}
\usepackage{epsfig,graphicx}
\usepackage{amsmath}
\usepackage{natbib}
\usepackage{comment}
\usepackage{url}
\usepackage{float}

\usepackage{siunitx}
\usepackage{booktabs}
\usepackage{multirow}
\usepackage{xcolor}
\usepackage{subfigure}
\usepackage{lineno}
\usepackage[normalem]{ulem}

\definecolor{maroon}{rgb}{0.760,0.118,0.337}

\definecolor{darkgreen}{rgb}{0.2,0.6,0.2}

\newcommand{\tardis}{\textsc{tardis}}

\newcommand\kms{~km~s$^{-1}$}

\shorttitle{SN~2014ad}
\shortauthors{Kwok et al.}

\submitjournal{\textit{The Astrophysical Journal}}
\received{2022 April 1}
\accepted{2022 August 11}
\begin{document}
\turnoffeditone

\title{Ultraviolet Spectroscopy and TARDIS Models of the Broad-lined Type-Ic Supernova 2014ad}

\correspondingauthor{Lindsey A. Kwok}
\email{lindsey.kwok@physics.rutgers.edu}

\author[0000-0003-3108-1328]{Lindsey A. Kwok}
\affiliation{Department of Physics \& Astronomy, Rutgers, the State University of New Jersey, Piscataway, NJ 08854, USA}

\author{Marc Williamson}
\affiliation{Department of Physics, New York University, New York, NY 10003, USA}

\author[0000-0001-8738-6011]{Saurabh W. Jha}
\affiliation{Department of Physics \& Astronomy, Rutgers, the State University of New Jersey, Piscataway, NJ 08854, USA}

\author{Maryam Modjaz}
\affiliation{Department of Physics, New York University, New York, NY 10003, USA}

\author[0000-0002-9830-3880]{Yssavo Camacho-Neves}
\affiliation{Department of Physics \& Astronomy, Rutgers, the State University of New Jersey, Piscataway, NJ 08854, USA}

\author{Ryan J. Foley}
\affiliation{Department of Astronomy and Astrophysics, University of California, Santa Cruz, CA 95064, USA}

\author{Peter Garnavich}
\affiliation{Department of Physics, University of Notre Dame, Notre Dame, IN 46556, USA}

\author[0000-0003-2611-7269]{Keiichi Maeda}
\affiliation{Department of Astronomy, Kyoto University, Kitashirakawa-Oiwake-cho, Sakyo-ku, Kyoto, 606-8502. Japan}

\author{Dan Milisavljevic}
\affiliation{Department of Physics and Astronomy, Purdue University, West Lafayette, IN 47907 USA}
\affiliation{Integrative Data Science Initiative, Purdue University, West Lafayette, IN 47907, USA}

\author{Viraj Pandya}
\altaffiliation{Hubble Fellow}
\affiliation{Department of Astronomy, Columbia University, New York, NY 10027, USA} 

\author{Mi Dai}
\affiliation{Department of Physics \& Astronomy, Rutgers, the State University of New Jersey, Piscataway, NJ 08854, USA}
\affiliation{Department of Physics \& Astronomy, Johns Hopkins University, Baltimore, MD 21218, USA}

\author[0000-0001-5807-7893]{Curtis McCully}
\affiliation{Las Cumbres Observatory, Goleta, CA 93117, USA}
\affiliation{Department of Physics, University of California, Santa Barbara, CA 93106, USA}

\author{Tyler Pritchard}
\affiliation{Department of Physics, New York University, New York, NY 10003, USA}

\author[0000-0002-8310-0829]{Jaladh Singhal}
\affiliation{TARDIS Collaboration}



\begin{abstract}
Few published ultraviolet (UV) spectra exist for stripped-envelope supernovae, and none to date for broad-lined Type Ic supernovae (SN Ic-bl). These objects have extremely high ejecta velocities and are the only supernova type directly linked to gamma-ray bursts (GRBs). Here we present two epochs of \textit{HST/STIS} spectra of the SN Ic-bl 2014ad, the first UV spectra for this class. We supplement this with 26 new epochs of ground-based optical spectra, augmenting a rich spectral time series. The UV spectra do not show strong features and are consistent with broadened versions of other SN Ic spectra observed in the UV. We measure Fe II 5169~\AA\ velocities and show that SN 2014ad has even higher ejecta velocities than most SNe Ic both with and without observed GRBs. We construct models of the SN 2014ad UV+optical spectra using \tardis, a 1D Monte-Carlo radiative-transfer spectral synthesis code. The models fit the data well at multiple epochs in the optical but underestimate the flux in the UV, likely due to simplifying assumptions. We find that high densities at high velocities are needed to reproduce the spectra, with $\sim$3~M$_\odot$ of material at $v >$ 22,000~\kms, assuming spherical symmetry. Our nebular line fits suggest a steep density profile at low velocities. Together, these results imply a higher total ejecta mass than estimated from previous light curve analysis and expected from theory. This may be reconciled by a flattening of the density profile at low velocity and extra emission near the center of the ejecta.
\end{abstract}



\section{Introduction \label{sec:intro}} 

Core-collapse supernovae (SNe), arising from the death of massive stars, are some of the most energetic cosmic explosions and serve to energize and chemically enrich their surroundings. When a massive ($\gtrsim$ 8 M$_\odot$) star loses part or all of its outer layers of hydrogen and helium prior to explosion, the resulting supernova (SN) is called a stripped-envelope SN \citep{Clocchiatti1997}. Type Ic supernovae (SNe Ic) are stripped-envelope SNe that display an absence of both hydrogen and helium features in their photospheric spectra \citep{Filippenko1993,Gal-Yam2017,Modjaz19}. The progenitors and causes of stripping for SNe Ic are not known definitively: they may result from Wolf-Rayet stars with powerful stellar winds that remove the hydrogen and helium envelopes \citep{Crowther2007, Smartt2009}; slightly less massive stars with a binary companion that removes the helium envelope through mass transfer \citep{Podsiadlowski1992, Yoon2015}; or homogeneous chemical evolution due to rotation that mixes the hydrogen internally in low metallicity stars \citep{Maeder1987} \citep[for more review, see][]{Langer2012}. Furthermore, the extent of stripping and the existence of hidden helium remains an open question for SNe Ic \citep{Filippenko1995, Dessart2012, Williamson2021}.

Broad-lined Type Ic supernovae (SNe Ic-bl) are a rare (4\% of stripped-envelope SNe; \citealt{Shivvers2017}) and extreme form of this class. Like SNe Ic, SNe Ic-bl spectra lack hydrogen and helium, but distinctively, they exhibit broad and blended spectral features, caused by Doppler shifting due to high photospheric velocities ($\sim$15,000--30,000\kms) of the ejecta \citep{mlb16}. On average, SNe Ic-bl also have higher luminosities and higher kinetic energies than regular SNe Ic \citep{Prentice2016,Cano_2017,Mazzali21}. A unique aspect of SNe Ic-bl is that they are the only SN type that has been directly associated with long-duration gamma-ray bursts \citep[GRBs; for reviews, see][]{wb2006,Modjaz11,Cano_2017}. However, not all SNe Ic-bl have an observed accompanying GRB. SNe Ic-bl associated with GRBs (SN-GRBs) tend to have even higher velocities and broader lines than SNe Ic-bl without GRBs \citep{Modjaz2016}. Efforts to understand this connection between SNe and GRBs have motivated searches, observations, and modeling of SNe Ic-bl.

The excess energy generated by SNe Ic-bl is thought to be supplied by a central engine such as an accreting black-hole or a magnetar. The central engine produces jets for the case when a GRB is observed with an associated SN Ic-bl. It is still uncertain whether SNe Ic-bl without GRBs are intrinsically different than SN-GRBs or if they are just seen off-axis from the jet. It is argued based on radio data constraints and modeling assumptions that not all SNe Ic-bl events can be explained as off-axis SN-GRBs \citep{Soderberg2006, Soderberg2010, Corsi_2016}. However, their assumptions mostly include high-energy GRBs and high-density ISM which are not very common amongst observed objects and typical GRB progenitor ISM. Thus, it is still plausible that SNe Ic-bl without observed GRBs harbored off-axis GRBs that were either of low energy or expanded in low-density media. Another potential explanation for a SN Ic-bl without an observed GRB is that it harbored a low-energy GRB with a jet, but the jet was choked via thick surrounding material or an additional progenitor envelope \citep{Margutti2014}. The passage of the jet would still accelerate the ejecta to high velocities (though perhaps not to velocities as high as for SN-GRBs, see \citealt{Modjaz2016}), but the jet would be smothered \citep{Cano_2017, lazzati2012unifying}. Indeed, \citet{Modjaz20} argue for both cases for their PTF SN Ic-bl sample that reside in low-metallicity galaxies, similar to those of SN-GRBs. 

Our understanding of SNe Ic-bl has been greatly enhanced by increased discovery rates from wide-field surveys and improved optical data coverage by networks of follow-up telescopes. However, the sample of ultraviolet (UV) data available for SNe Ic and Ic-bl remains sparse, due to the need for space-based observations. Current UV spectroscopy of SNe Ic includes only one published spectrum of PTF\,12gzk \citep{Ben-Ami2012} (full UV spectroscopic analysis including other epochs has yet to be published) and unpublished spectra of SN~1994I. To our knowledge, no UV spectra of a SN Ic-bl has been previously published. Most efforts to obtain UV spectra of SNe have been targeted at other supernova classes \citep[e.g.][]{Foley2020,Miller2021, Vasylyev2022}.

UV emission is strongest at early times when the ejecta are hot, typically when the photosphere is located in the high velocity layers of the ejecta. At these early times the optically thick UV wavelengths probe the less radioactively processed outer layers of the SN. Thus, the UV is sensitive to information about the explosion physics, progenitors, and environments of the SN \citep{Bufano2009}. Additional data, especially spectra, in the UV could improve our current models and deepen our understanding of the differences in progenitor systems that give rise to SNe Ic, SNe Ic-bl, and SN-GRBs.

In this work we present UV data of SN~2014ad. SN~2014ad was discovered by \cite{Howerton2014} on March 12.4, 2014 (MJD 56728.4) in public images from the Catalina Real-Time Transient Survey \citep{Djorgovski2011} at $\alpha$ = 11$^\mathrm{h}$57$^\mathrm{m}$44$\fs$44, $\delta = -10\arcdeg 10\arcmin 15\farcs7$. It is located in PGC 37625 (Markarian 1309) with a redshift $z=0.005723$ \citep{Wong2006} and a distance of 28 $\pm$ 2 Mpc, estimated using the redshift, correcting for Local Group infall onto the Virgo Cluster (via the NASA Extragalactic Database, NED). Originally thought to be a peculiar type Ia supernova, further spectroscopy led to its classification as a broad-line type Ic SN \citep{Howerton2014}. 

Several previous studies in the literature characterize many of SN~2014ad's properties. \cite{Sahu2018} present optical imaging and spectroscopy of SN~2014ad and \cite{Stevance2017} present optical spectropolarimetry. From their analysis \citet{Sahu2018} find a total reddening (Milky Way $+$ host galaxy) to the supernova of $E(B-V)_{\text{total}} = 0.22 \pm 0.06$ mag; we adopt this value here.

These studies found that SN~2014ad had an ejecta velocity of $\sim$ 30,000\kms\ at early times, kinetic energy of $(1.0\pm0.3)\times10^{52}$ erg, ejecta mass of $3.3\pm0.8$ M$_\odot$, and host galaxy metallicity of $\sim 0.5$ Z$_\odot$. \cite{Sahu2018} suggest a massive star progenitor with zero-age main-sequence (ZAMS) mass M$_{\text{ZAMS}} \gtrsim 20$ M$_\odot$ and \cite{Stevance2017} favor a mildly aspherical outer ejecta with a nearly spherical inner ejecta. No GRB was discovered to accompany SN~2014ad and the deep radio and X-ray constraints presented in \citet{Marongiu2019} suggest that while an off-axis GRB cannot be ruled out, it would require high jet collimation (e.g. $\theta_{obs}\gtrsim30^{\circ}$, $\theta_j=5^{\circ}$) in a very low-density circumstellar medium (CSM) environment. The spectropolarimetric data indicate that the axial symmetry at early times is not inconsistent with a jet, but the spherical inner ejecta at late times is difficult to reconcile with the presence of a highly collimated jet \citep{Stevance2017}.

Here we present two early-time epochs of UV spectroscopy of SN~2014ad from the \textit{Hubble Space Telescope} (\textit{HST}). We also present a well-sampled spectroscopic time series showing the evolution of SN~2014ad with 22 new epochs of optical spectra from the Southern African Large Telescope (SALT). In \autoref{sec:observations} we describe our observations and data reduction, in \autoref{sec:spec_analysis} we present analysis of the optical and UV spectra, in \autoref{sec:models} we present spectral modeling results from the radiative transfer code \tardis, and in \autoref{sec:nebular_analysis} we present analysis of the nebular spectra. We discuss our results and conclude in \autoref{sec:discussion}.

\begin{figure}[t]
    \centering
    \includegraphics[width=0.48\textwidth]{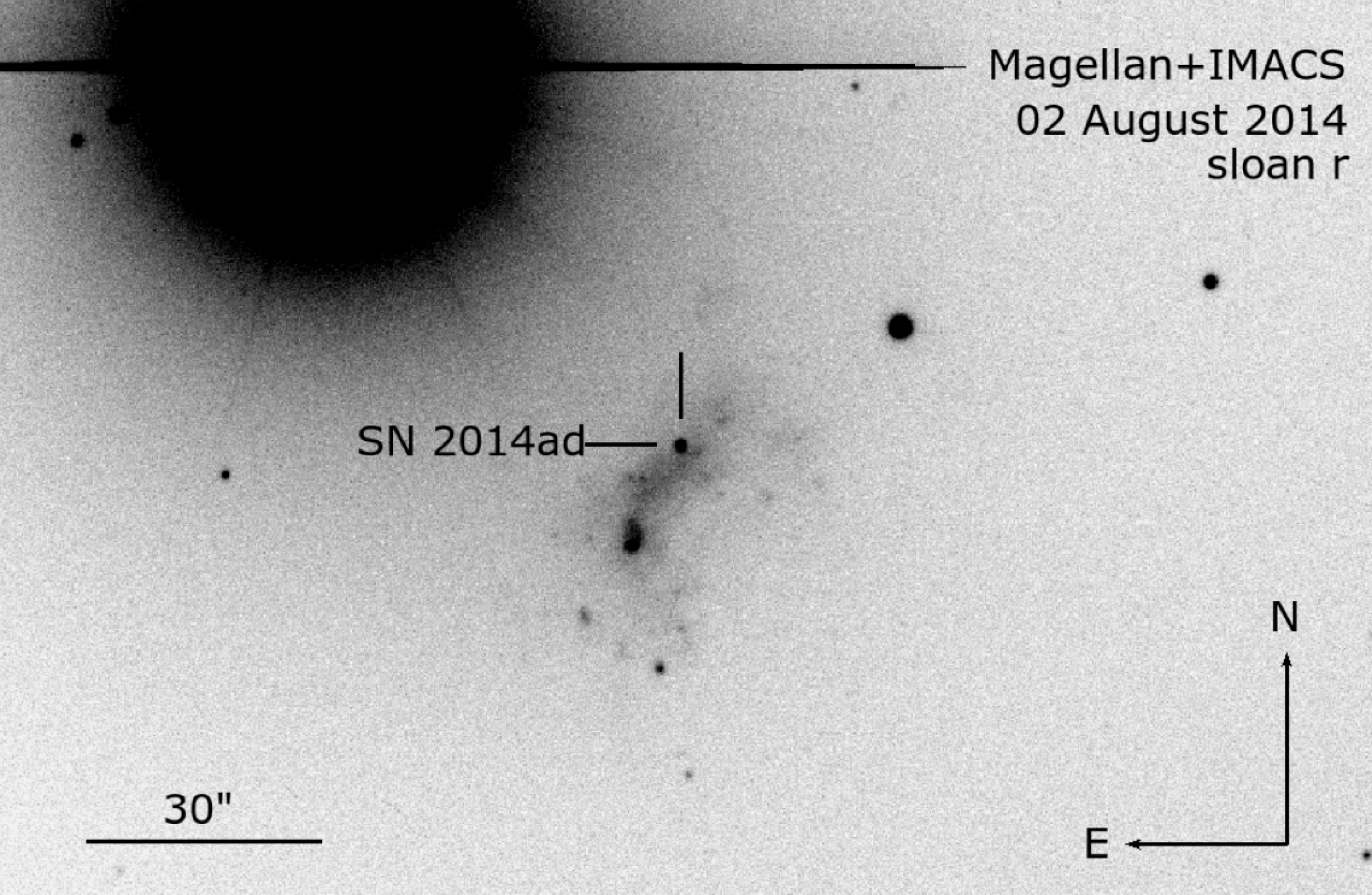}
    \caption{Finding chart for the 2 Aug 2014 spectrum of SN~2014ad from Magellan+IMACS. SN~2014ad was difficult to observe due to the nearby \textit{V}$=$7.4~mag star.}
    \label{fig:fchart}
\end{figure}

\begin{figure*}[h!]
    \centering
    \includegraphics[width=\textwidth]{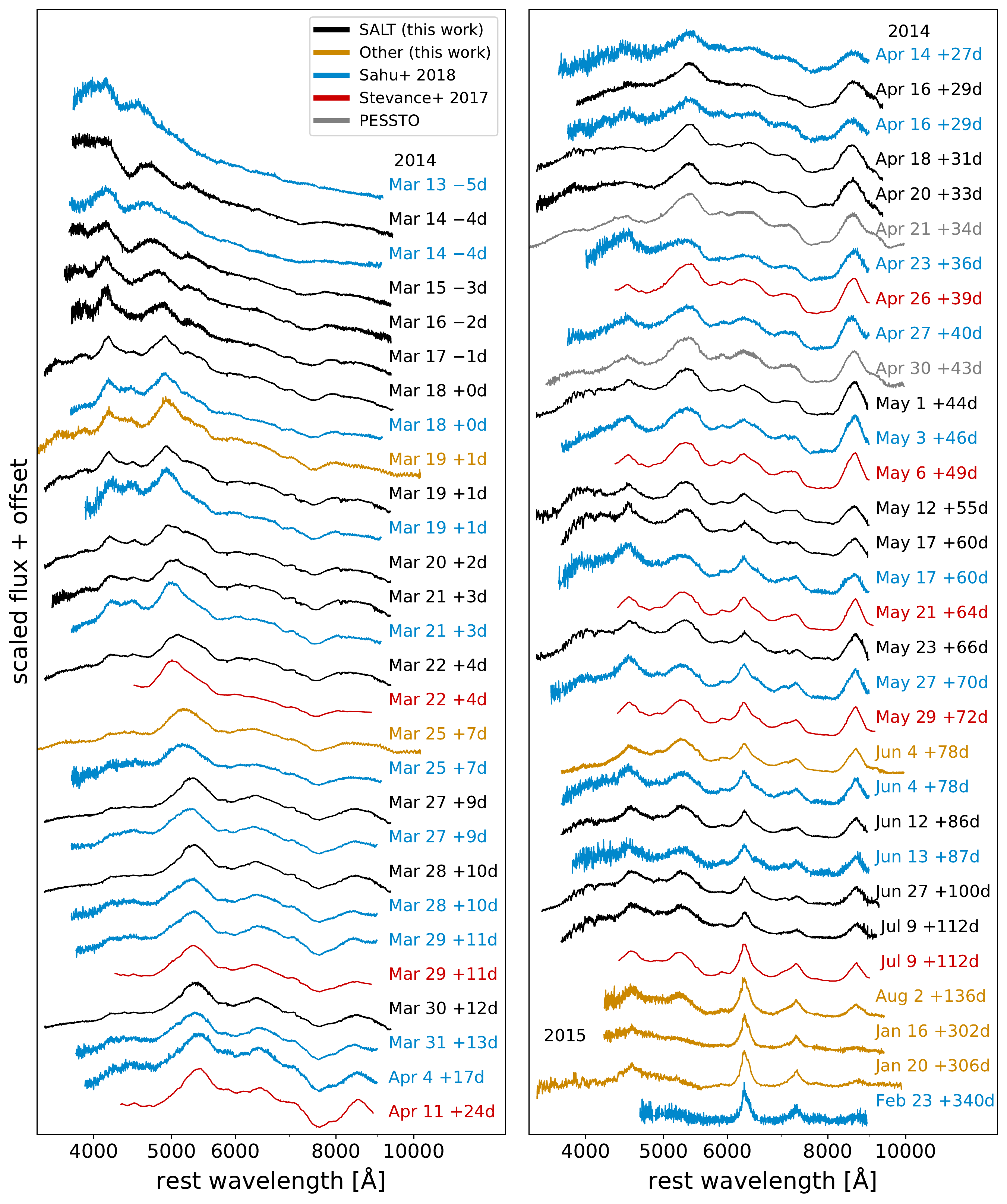}
    \caption{Optical spectral timeseries of SN~2014ad. We present 22 epochs from SALT (black), two epochs from \textit{HST} (yellow; 2014-Mar-19, 2014-Mar-25), two epochs from LBT (yellow; 2014-Jun-04, 2015-Jan-20), and two epochs from Magellan (yellow; 2014-Aug-02, 2015-Jan-16). We display published spectra from \citet[][blue]{Sahu2018} and \citet[][red]{Stevance2017} as well as two unpublished epochs (gray) from PESSTO \citep{PESSTO} via WISeREP \citep{WISEREP}. The wavelength axis is redshift-corrected to the host galaxy rest-frame. The fluxes have been corrected for total reddening and are scaled and vertically shifted for clarity. Cosmic rays, narrow host-galaxy lines and telluric absorption have been removed in all spectra. The phase is given in rest-frame days relative to \textit{B} maximum light \citep[MJD 56735.11;][]{Sahu2018}. \label{fig:all_spec}}
\end{figure*}

\section{Observations \label{sec:observations}} 

\begin{figure*}[t]
    \centering
    \includegraphics[width=0.95\textwidth]{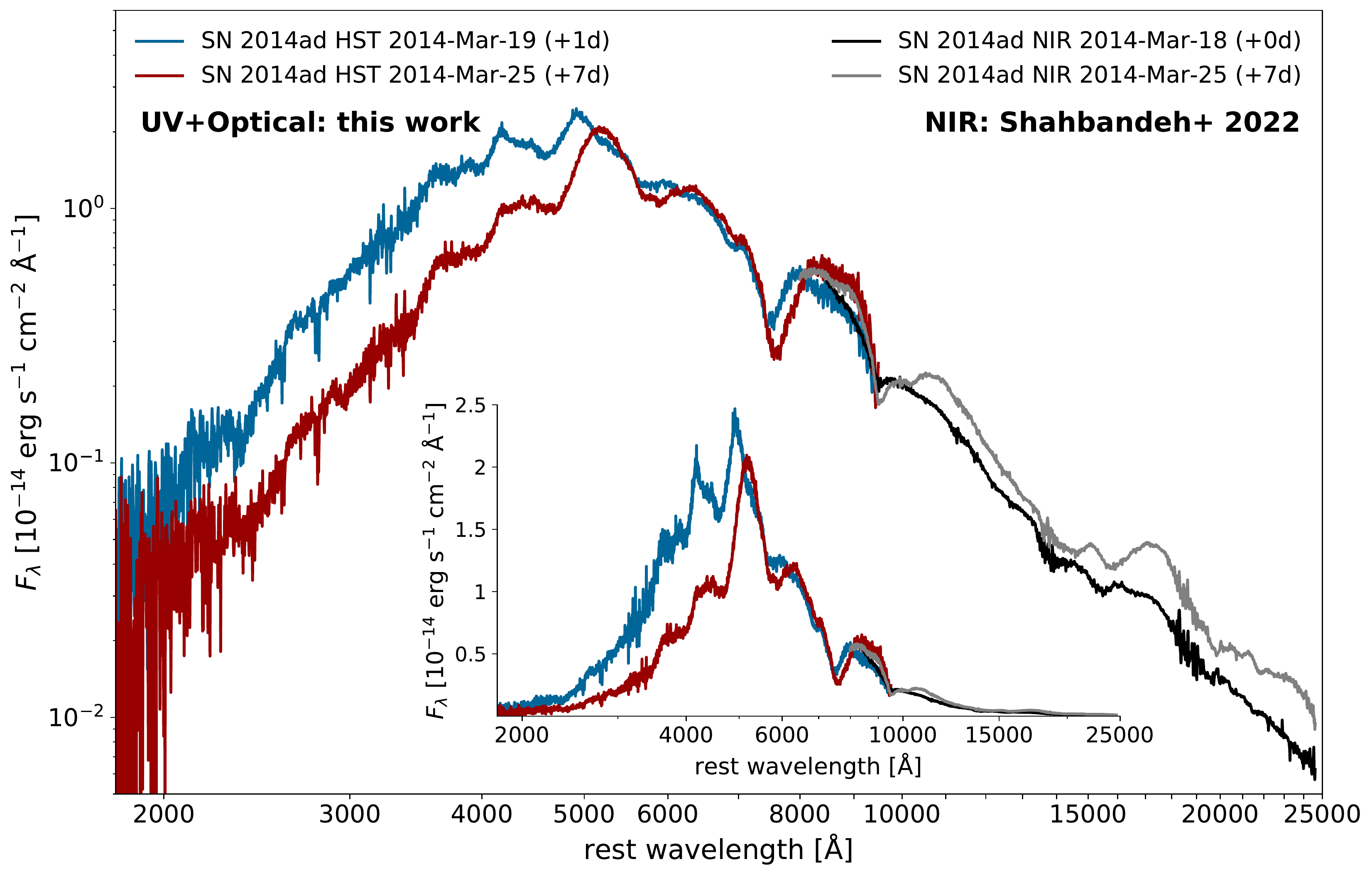}
    \caption{Complete ultraviolet (UV) to near-infrared (NIR) spectra of SN~2014ad at two epochs near maximum light. We present \textit{HST/STIS} spectroscopy of SN~2014ad covering the UV and optical wavelength ranges. The 2014-Mar-19 \textit{HST} spectrum corresponds to a phase of $+1$d (blue) and the 2014-Mar-25 \textit{HST} spectrum corresponds to a phase of $+7$d (red) past \textit{B} maximum light. NIR spectra of SN~2014ad from \citet{Shahbandeh_2022} at closely matching epochs of $+0$d (black) and $+7$d (grey) are plotted alongside the UV$+$Optical spectra to show the complete composite spectrum. The spectra have been plotted on a log-log scale to emphasize the UV and NIR data, while the inset shows the data on a log-linear scale. \label{fig:HST_spec}}
\end{figure*}

\subsection{SALT/RSS Optical Spectroscopy \label{sec:SALT}} 

We present 22 epochs of SALT optical spectra of SN~2014ad in \autoref{fig:all_spec}, taken between UT 2014 March 13 and 2014 July 9. The data were obtained with the Robert Stobie Spectrograph \citep[RSS; ][]{RSS} using a 1\farcs5 longslit and the PG0900 grism. The spectra were reduced using a custom pipeline based on standard Pyraf \citep{Pyraf} spectral reduction routines and the PySALT package \citep{PySALT}. 

We redshift-correct the spectra to the host-galaxy rest frame and remove cosmic rays, host galaxy lines, and telluric lines. We scale the spectra to match the \textit{V}-band photometry from \cite{Sahu2018} and correct for total reddening using the \citet{Fitzpatrick1999} dust extinction model with $R_V = 3.1$ and $E(B-V) = 0.22$ mag as described above. In some cases the spectra did not fully span multiple filters, while in others, scaling the spectra with multi-band photometry produced unacceptably large distortions to the continuum shapes of spectra near in time, so we chose to scale to the \textit{V}-band only. In a few late-time epochs, there is a discrepancy in the continuum shape of the SALT data that may have arisen from galaxy contamination or flux calibration difficulties because standard stars are not observed at the same time as the supernova data. In those cases we adjusted the continuum with a low-order polynomial to match the spectral shape of other spectra at nearby epochs. In this work we define the spectral phase in rest-frame days relative to \textit{B}-band maximum, 2014 March 18 \citep[MJD 56735.11; ][]{Sahu2018}. Further details of the spectroscopic observations can be found in \autoref{tab:spec}.

\subsection{\textit{HST/STIS} UV+Optical Spectroscopy \label{sec:HST}} 

We present the first published UV spectra of a SN Ic-bl in \autoref{fig:HST_spec}. SN~2014ad was observed by \textit{HST} using the \textit{STIS} spectrograph on 2014 March 19.5 and 25.4.  The first and second visits consisted of two and four orbits, respectively; however, telescope problems resulted in the second epoch losing data for two exposures. The observations in each visit were obtained with three different gratings, all with the $52\arcsec \times$ 0\farcs2 slit.  The first (second) epoch had a total exposure of 3478~s (4946~s) utilizing the near-UV MAMA detector and the G230L grating. The first epoch obtained 400~s of explosion with both the CCD/G430L and CCD/G750L setups, respectively, while the second epoch obtained 700 and 400~s of exposure with the CCD/G430L and CCD/G750L setups, respectively.  The three setups yield a combined wavelength range of 1615--10230~\AA.

The data were reduced using the standard \textit{HST} Space Telescope Science Data Analysis System (STSDAS) routines to bias subtract, flat-field, extract, wavelength-calibrate, and flux-calibrate each SN spectrum \citep{Foley2012, Foley2014, Foley2016}. We remove cosmic rays, rebin in 2 \AA\ bins, and combine the MUV G230L and NUV/optical G430L spectra. As with the optical data, we correct the UV spectra for redshift and total reddening. The log of spectroscopic observations is given in \autoref{tab:spec}.

In \autoref{fig:HST_spec}, we also show two epochs of near-infrared (NIR) spectra for SN~2014ad from \citet{Shahbandeh_2022} on 2014 March 18 ($+0$d) and 2014 March 25 ($+7$d) that closely match the phase of our two epochs of \textit{HST} UV$+$Optical spectra on 2014 March 19 ($+1$d) and 2014 March 25 ($+7$d). Combined, these data give the composite UV + Optical + NIR spectra of SN~2014ad for two epochs near maximum light.

\subsection{Nebular Optical Spectroscopy \label{sec:nebular}} 

We obtained five spectroscopic exposures of SN~2014ad with the Large Binocular Telescope (LBT) using the Multi-Object Dual Spectrographs \citep[MODS;][]{pogge12} on 2015 January 20.4 (UT), 306 rest-frame days after maximum light. The exposure time totaled 3000s. A 1.0 arcsec wide slit was employed under good conditions with the seeing steady at about 1.0 arcsec. Using the dual grating mode of MODS1, the red spectrograph covered the wavelength range of 5600--10100~\AA\ at a resolution of 240~\kms\ (FWHM).  The blue side of the spectrograph typically covers 3200~\AA\ to 5600~\AA, but no trace of the supernova was evident shortward of 3800~\AA.

The images were averaged using a min/max rejection algorithm to remove cosmic ray events. The blue side was wavelength calibrated using a neon emission line lamp while an argon lamp was employed for the red spectrum. The full wavelength range was flux calibrated using a spectrum of the spectrophotometric standard Feige~34 obtained on the same night as the SN~2014ad spectra.

Two epochs of optical spectra were also obtained with the 6.5~m Magellan telescope at Las Campanas Observatory on 2014 August 2.0 and 2015 January 16.3. The Inamori Magellan Areal Camera and Spectrograph (IMACS; \citealt{Dressler11}) was used with the $f/2$ camera in combination with the Mosaic3 array of eight thinned 2K $\times$ 2K $\times$ $15\,\mu$m E2V CCDs. A 300 line mm$^{-1}$ grism and a 0\farcs9 long slit was used. Exposures of $2 \times 1200$\,s and $3 \times 1800$\,s were obtained and averaged at each epoch, respectively. Resulting spectra have an effective wavelength range of 4230–-9400 \AA, with a dispersion of 1.3 \AA\ pixel$^{-1}$ and FWHM resolution of 4 \AA\ (measured at 6000 \AA). 

Standard procedures to bias-correct, flat-field, and flux-calibrate the data were followed using IRAF \citep{IRAF1, IRAF2}. LA-Cosmic \citep{vanDokkum01} was used to remove cosmic rays in individual images. Some cosmetic defects introduced from hot pixels and imperfect background subtraction have been manually removed. Spectrophotometric standards LTT7379, Feige 110, LTT1788, and LTT3864 were observed and used for absolute flux calibration, which is believed to be accurate to within 30\%.

\begin{figure}
    \centering
    \includegraphics[width=0.48\textwidth]{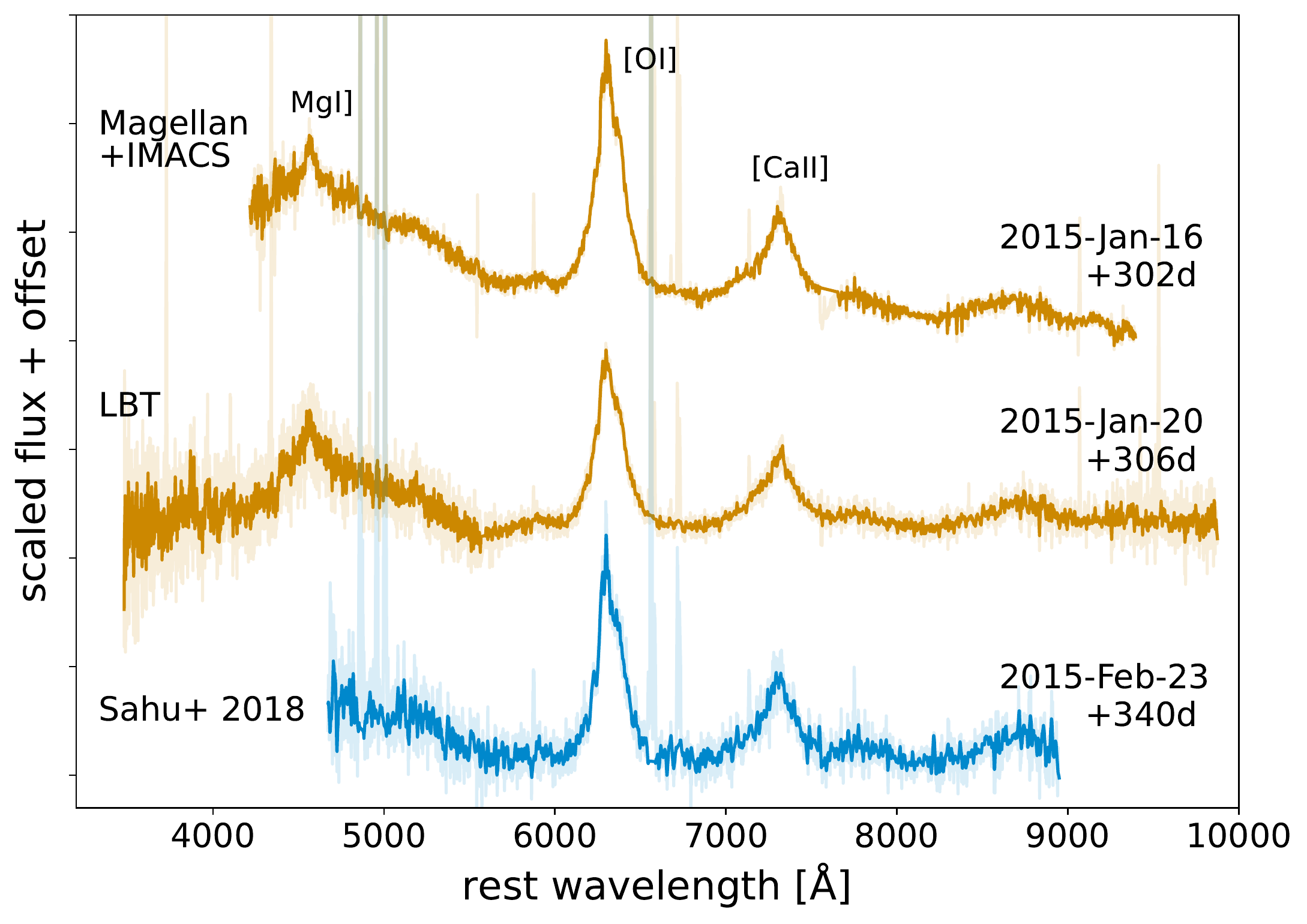}
    \caption{Nebular spectra of SN~2014ad showing our LBT and Magellan data, along with the late-time spectrum from \citet{Sahu2018}. The wavelength axis is redshift-corrected to the host galaxy rest-frame. The fluxes have been corrected for total reddening, and scaled and vertically shifted for clarity. Narrow lines and telluric absorption features were manually removed and the spectra were gently smoothed; the uncorrected spectra are shown as transparent lines. Phase is given in rest-frame days relative to \textit{B}-maximum \citep[MJD 56735.11;][]{Sahu2018} Spectra are colored to match \autoref{fig:all_spec}.  \label{fig:nebular}}
\end{figure}

\section{Spectral Analysis \label{sec:spec_analysis}} 

\subsection{Optical Spectral Evolution \label{sec:optical}} 

SN~2014ad has a rich optical spectral time series ranging from $-5$d to $+112$d relative to \textit{B}-maximum. \autoref{fig:all_spec} displays the compilation of spectra from the existing literature \citep{Sahu2018, Stevance2017}, from this work (SALT, \textit{HST}, LBT, and Magellan), and two unpublished PESSTO \citep{PESSTO} spectra available on WISeREP \citep{WISEREP}. All of the spectra have been corrected for redshift and total reddening. The data from different sources exhibit good agreement with each other. SN~2014ad's spectra exhibit broad and blended features and lack clear hydrogen or helium lines; hence its classification as a Ic-bl SN. 

At early times, before and near maximum light, the spectra are dominated by a blue continuum with broad absorption features due to high temperatures and velocities. \citet{Sahu2018} identify distinct broad absorption from Ca II ($\sim 3800$ \AA), Mg II ($\sim 4400$ \AA), and Fe II ($\sim 4800$ \AA). The absorption feature at $\sim 7300$ \AA\ is heavily blended O I and the Ca II NIR triplet; it suggests the presence of significant material at velocities $>30,000$ \kms\ \citep{Sahu2018, Mazzali2002}. The emission-like features at 4000 \AA\ and 4600 \AA\ are from regions of low opacity \citep{Sahu2018, Iwamoto1998, Mazzali2000}. SN~2014ad's features are even more blueshifted than other similar SN Ic-bl \citep{Sahu2018}, indicating exceptionally high velocities (see Section \ref{sec:FeII_vels}). 

The spectra evolve significantly after maximum light. As the SN expands and temperatures decline, the continuum shape shifts redder and prominent absorption and emission features emerge. The dominant features arise from Fe II, Si II, O I, and Ca II. The spectra evolve from mostly absorption to mostly emission features as the SN approaches the nebular phase \citep{Sahu2018}. All of the spectral features stay broad, indicating that the ejecta maintain high velocities. A comprehensive analysis of SN~2014ad's spectral evolution in the pre-maximum, maximum, and post-maximum phase is given by \citet{Sahu2018}. 

We present two additional epochs of optical nebular data for SN~2014ad, shown in \autoref{fig:nebular}. The spectra agree closely with the nebular spectrum from \citet{Sahu2018} which was taken about a month later. As discussed by \citet{Sahu2018}, the nebular spectra are dominated by [O I]~6300,~6364~\AA\ and [Ca II]~7291,~7324~\AA. Our nebular spectra additionally reveal a weak Mg I]~4571~\AA\ line feature, commonly observed in late-time spectra, and a fairly strong broad [Fe II] blend around 5200~\AA. \cite{Sahu2018} find a [O I]$/$[Ca II] line ratio of 1.54, which is comparable to Ic-bl SN~1998bw (1.7) and SN~2002ap (2). This line ratio is $>$ 1, implying a large main-sequence progenitor mass \citep{Nomoto2006, Maeda2007}. We further analyze the nebular spectra in \autoref{sec:nebular_analysis}.

\subsubsection{Fe II Velocity Analysis \label{sec:FeII_vels}}

SNe Ic-bl spectra are characterized by their highly blue-shifted, blended features due to extreme ejecta velocities ($\sim$15,000--30,000\kms). Some SNe Ic-bl have been observed in conjunction with long-duration GRBs, and those SN-GRBs exhibit systematically higher absorption velocities than their GRB-less Ic-bl counterparts \citep{Modjaz2016}. It is an open question whether GRB-less SNe Ic-bl are caused by choked jets \citep{lazzati2012unifying}, off-axis GRBs \citep{Barnes_2018}, on-axis-unobserved GRBs \citep[e.g., SN~2020bvc;][]{rho2020bvc}, or some combination of these scenarios. To investigate which of these scenarios may be consistent with SN~2014ad, we measure its absorption velocities.

\begin{figure}[t]
    \centering
    \includegraphics[width=0.48\textwidth]{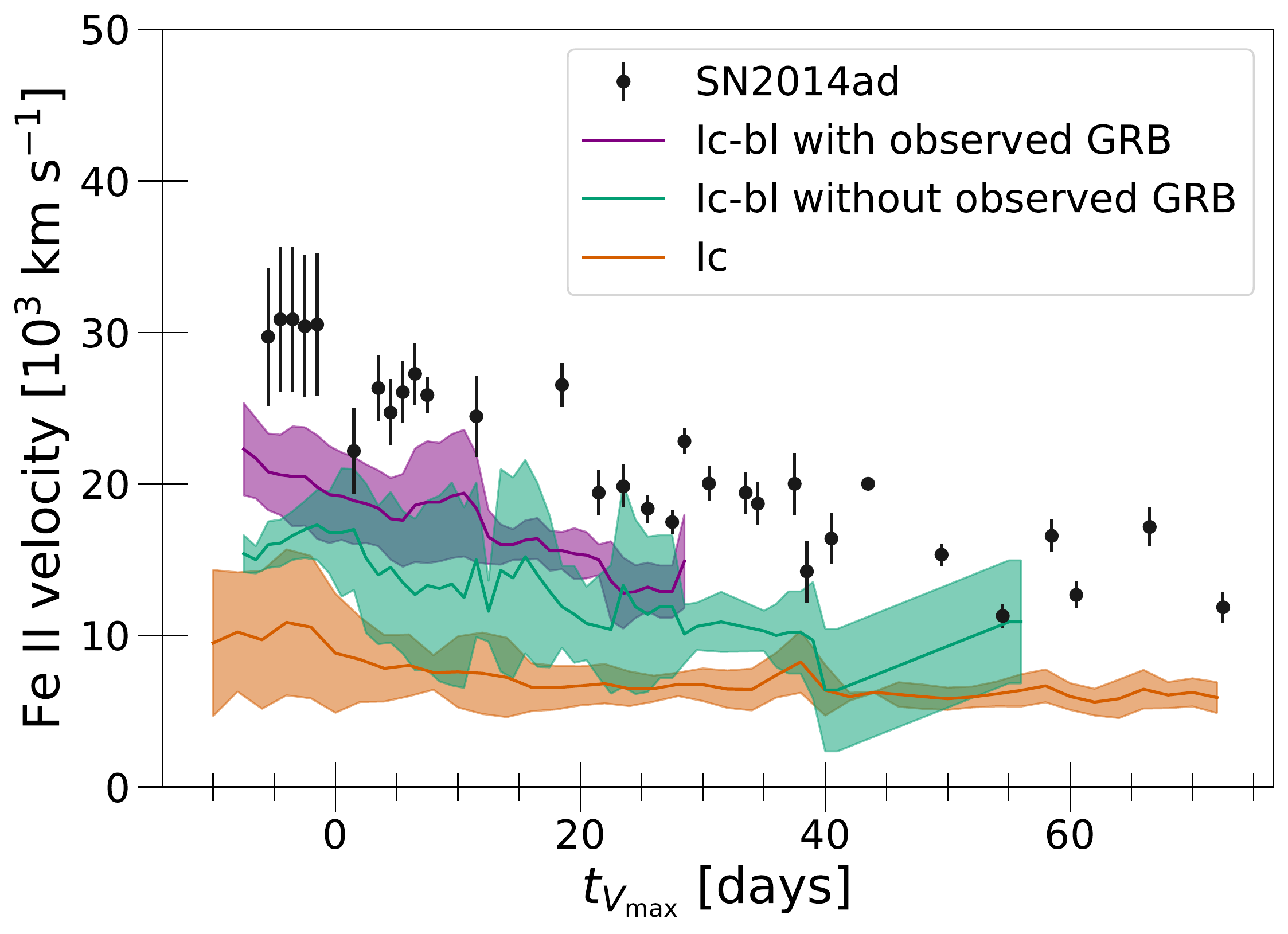}
    \caption{Fe II 5169 \AA\ velocities of SN~2014ad evolving over time calculated using the Monte Carlo technique from \citet{Modjaz2016}. SN~2014ad Fe II velocities (black) are compared to a sample of Ic (orange), Ic-bl without observed GRBs (teal), and SN-GRBs (purple). SN~2014ad consistently shows high ejecta velocities, exceeding those of typical Ic-bl, even those associated with GRBs. Note that time is given relative to \textit{V}$_{\mathrm{max}}$ which for SN~2014ad occurred at $\sim$5.5~days after \textit{B}$_{\mathrm{max}}$. \autoref{tab:FeVels} lists the Fe II velocities for SN~2014ad and their corresponding phases with respect to both \textit{B}$_{\mathrm{max}}$ and \textit{V}$_{\mathrm{max}}$.  \label{fig:FeVels}}
\end{figure}

Figure \ref{fig:FeVels} shows the temporal evolution of SN~2014ad's Fe II 5169 \AA\ line velocities compared to the velocity evolution of samples of normal SNe Ic, SNe Ic-bl without observed GRBs, and SN-GRBs. The Fe II expansion velocities are calculated using the Monte Carlo technique developed by \citet{Modjaz2016} which incorporates the effects of line broadening, and spectra are smoothed using a Fourier method \citep{Liu2016,williamson_2019}. There is a systematic underestimation of the uncertainties at later phases due to fewer SNe spectra contributing to the templates, artificially implying higher precision in measuring the lineshifts relative to the template sample. The scatter in SN~2014ad Fe II velocity (i.e. non-monotonicity) may be attributed to the underestimated uncertainties at later times and differences in how the spectra from different facilities were flux calibrated and reduced.

We find that SN~2014ad exhibits extremely high early velocities (consistent with the analysis from \citealt{Sahu2018}). In fact, the velocity evolution in SN~2014ad is systematically higher than both the samples of SNe Ic-bl without observed GRBs and SN-GRBs from \citet{Modjaz2016}. SN~2014ad also compares amongst the highest velocity objects in the iPTF/PTF SN Ic-bl sample from \citet{Taddia2019} (see their Figure 15), which was analyzed using the same method from \citet{Modjaz2016}. Our results would indicate that SN~2014ad may be more consistent with an on-axis-unobserved GRB scenario, such as that discussed by \citet{rho2020bvc} and \citet{Ho2020a} for the similar high velocity SN Ic-bl 2020bvc, although see \citet{Izzo2020} for an alternate explanation of SN~2020bvc as an off-axis GRB. However, for SN~2014ad this on-axis jet possibility is essentially ruled out by strong limits from X-ray and radio non-detections established by \citet{Marongiu2019}. Nevertheless, it is still possible that SN 2014ad may have harbored an off-axis low-energy jet, which the radio limits do not exclude \citep{Marongiu2019}. Indeed, off-axis GRBs can easily produce high-velocity ejecta that gives rise to spectra resembling those of SNe Ic-bl, as shown by \citet{Barnes_2018}. Future work on modeling the passage of off-axis vs. choked jets in the envelopes of exploding massive stars, coupled with radiative transfer calculations are needed to explain the high ejecta velocities in SN~2014ad given the lack of an associated on-axis GRB.

\begin{figure*}
    \centering
    \includegraphics[width=\textwidth]{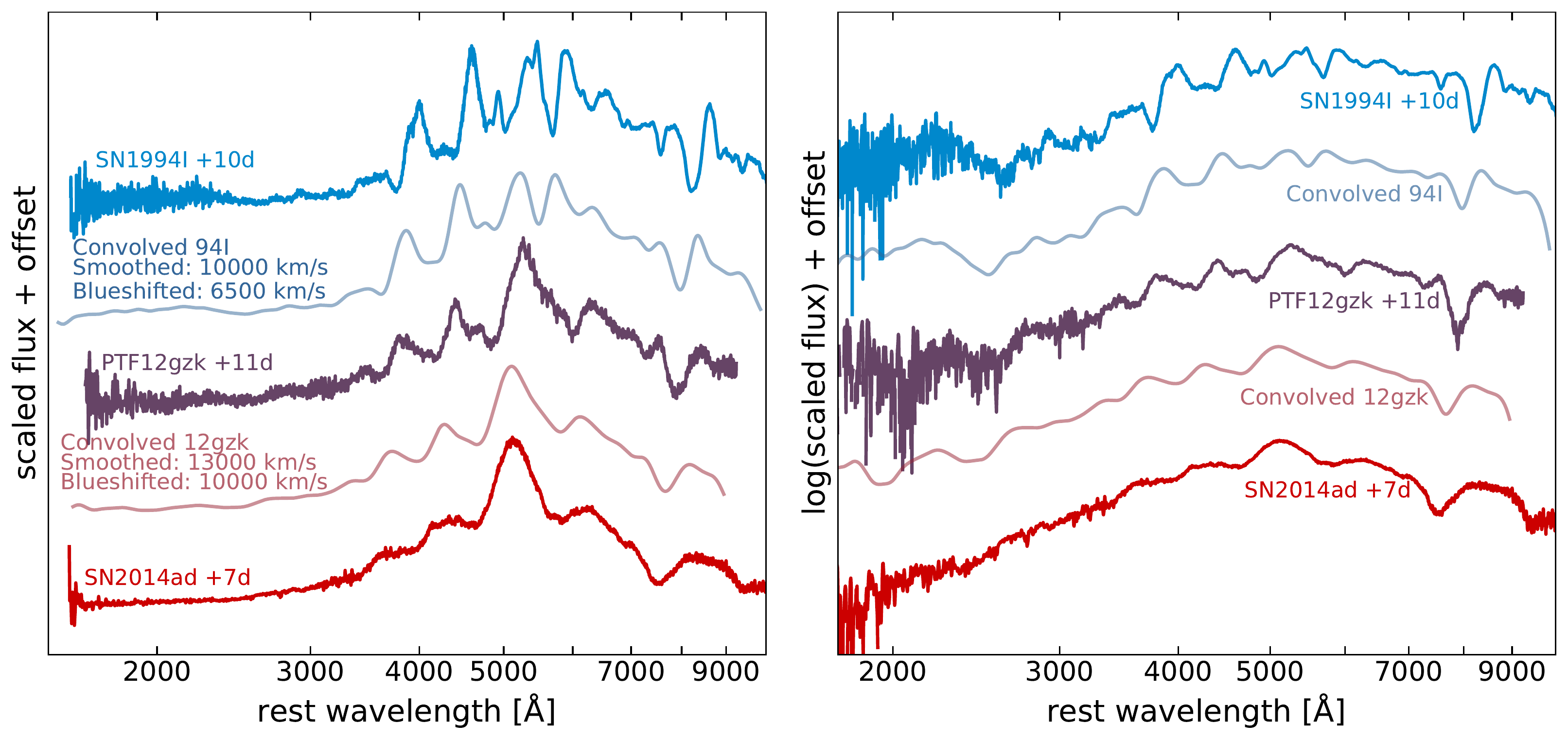}
    \caption{Comparison between the observed UV spectrum of SN~1994I on 19 Apr 1994 corresponding to phase $+$10d, observed UV PTF\,12gzk spectrum from 15 Aug 2012 corresponding to phase $+$11d, and the observed UV SN~2014ad spectrum from 25 Mar 2014 corresponding to phase $+$7d. SN~1994I was smoothed and blueshifted to match PTF\,12gzk and then PTF\,12gzk was smoothed and blueshifted to match SN~2014ad. The smoothing is given as the FWHM of the Gaussian convolution kernel (in velocity space). The right panel is the same as the left panel, but plotted on a logarithmic scale to better show the UV flux. \label{fig:convolution}}
\end{figure*}

\subsection{UV+Optical Analysis \label{sec:UV+opt}} 

We can place SN~2014ad in context of other extreme, stripped-envelope SNe through comparisons to other SN Ic, Ic-bl and SN-GRBs. We begin by focusing on the UV spectra, comparing SN~2014ad to the closest comparison objects with UV spectra: SN~1994I and PTF\,12gzk. SN~1994I has canonically been thought of as a typical Ic, but \citet{Modjaz2016} suggest it is not representative of the average SN~Ic. SN~1994I has an \textit{HST} FOS/RD UV spectrum in the MAST archive (G190H+G270H+G400H; program ID 5623, PI: Kirshner), observed on 1994 Apr 19 at $+10$ days past maximum light \citep[shown also by][]{Chornock:2013}. PTF\,12gzk is a peculiar SN Ic with line widths similar to those of normal SN Ic, but with much larger absorption velocities (blueshifts), similar to those of SN Ic-bl \citep{Modjaz2016}. A near-maximum light UV spectrum of PTF\,12gzk is presented by \citet{Ben-Ami2012}; here we analyze a later-phase \textit{HST/STIS} NUV-MAMA+CCD (G230L+G430L+G750L) spectrum observed on 2012 Aug 15, $+11$ days past maximum light, obtained through the MAST archive (program ID 12530; PI: Filippenko).

The UV spectra for these three SN~Ic, all at roughly similar phase, are shown in \autoref{fig:convolution}. While there are clearly some similarities between these distinct objects, the differing line velocities makes it hard to directly compare them. Thus, we follow the approach of \citet{Modjaz2016}, smoothing the SN Ic spectra by convolving with a Gaussian and applying a blueshift. This technique roughly simulates the line broadening and line shifting of the spectrum that would occur if a SN Ic had higher velocities like those of a SN Ic-bl. As shown in \autoref{fig:convolution}, we find that the spectra transition nicely from SN~1994I to PTF\,12gzk to SN~2014ad: smoothed versions of lower-velocity SN~Ic do a reasonable job of matching the observations of higher-velocity SN Ic in these cases. 

This result is in accord with the findings of \citet{Modjaz2016}, but they note that in detail, SN Ic-bl are more than just higher velocity versions of regular SN Ic. Similarly, for the UV+optical spectra that we present, there are also important differences to note. For instance, while \autoref{fig:convolution} shows that the smoothed and blueshifted version of SN~1994I produces a spectrum similar to that of PTF\,12gzk, there is a strong Na I absorption feature near 5800 \AA\ in SN~1994I that is not present in PTF\,12gzk or SN~2014ad. Smoothing and blueshifting PTF\,12gzk's spectrum produces a closer match to SN~2014ad than convolving SN~1994I's spectrum directly to SN~2014ad velocities. In our spectral convolution comparison, PTF\,12gzk looks like a transition object between a SN Ic and a SN Ic-bl. 

Shown in \autoref{fig:HST_spec}, the UV spectra of SN~2014ad from \textit{HST} exhibit low flux levels and have few distinct SN features. In both epochs a shoulder feature can be seen near 2600 \AA, with a possible faint feature around 2300 \AA. While the spectra of SN~1994I and PTF\,12gzk spectra are quite noisy below $\sim$2400 \AA, they display similar shapes and features to each other and to SN~2014ad in the UV. The Ic UV spectral features are slightly more distinct; however, these features mostly broaden and blend out when convolved to higher velocities. \citet{Modjaz2016} found that the convolution of the mean SN Ic spectrum reproduces the mean SN Ic-bl spectrum better at longer wavelengths ($>$5000 \AA) than at shorter wavelengths.

Modeling work has suggested that the UV spectra of SNe Ic should be sensitive to progenitor metallicity \citep{Sauer2006}, given the many Fe-group element transitions in the UV; thus, it is expected that SNe with different progenitor metallicities would exhibit different UV behavior. However, we find that the UV spectra of SNe Ic-bl and those of SNe Ic after convolution are very similar to each other despite a large range in metallicities. As probed by environmental HII region oxygen abundance (in the PP04-O3N2 scale; \citealt{Pettini04}), PTF\,12gzk has a metallicity of $\log\mathrm{(O/H)}+12 \simeq$ 8.0 \citep{Modjaz20}, while SN~2014ad has $\log\mathrm{(O/H)} +12 \simeq$ 8.4 \citep{Sahu2018}, and SN~1994I is probably solar or super-solar. This corresponds to almost a factor of 10 in metallicity. Our results may imply that the smoothing due to high velocities, especially those of SN~2014ad, smears out any strong differences in the UV spectra due to progenitor metallicities.

\subsection{UV + Optical + NIR Composite Spectrum \label{sec:uv+opt+nir}}

We can combine the analysis above with the results from \citet{Shahbandeh_2022} to understand the UV + Optical + NIR composite spectrum. NIR spectra are particularly interesting for stripped-envelope SNe because two strong He absorption features in the NIR can probe the amount of helium leftover in the explosion. The He I $\lambda$1~$\mu$m absorption can be contaminated by a mix of C I $\lambda$1.0693~$\mu$m or Mg II $\lambda$1.0927~$\mu$m lines, so the uncontaminated He I $\lambda$2.0581~$\mu$m absorption line is a better indicator of helium in the supernova ejecta \citep{Dessart2015, Williamson2021, Shahbandeh_2022}. As noted in the figure caption of Figure C2 \& C3 from \citet{Shahbandeh_2022}, the He I absorption features of SN~2014ad in both the 1~$\mu$m and 2~$\mu$m regions are not strong enough to be fit in the earliest two NIR spectra ($+0$d and $+7$d). Evidence of broad He absorption is present in the NIR spectra of SN~2014ad from \citet{Shahbandeh_2022} at later epochs ($+35$d and $+79$d), though the pEW measurements in the 2 $\mu$m region are still weaker than all of the objects in the He-rich (SN IIb/Ib) group at similar phase. The NIR spectra of SN~2014ad published by \citet{Shahbandeh_2022} reveals that helium does not appear in SN~2014ad at early times, but it does appear later on. \citet{Sahu2018} showed that inclusion of He I marginally improves their \textsc{SYN++} synthetic fits of SN~2014ad around 5400 \AA\ at early phases of $+0$d and $+3$d, but they caution that confirmation of the presence of helium would require NIR spectra and detailed spectral modeling. We further discuss the question of helium in the ejecta of SN~2014ad in \autoref{sec:results}.

\begin{figure*}
    \centering
    \includegraphics[width=0.82\textwidth]{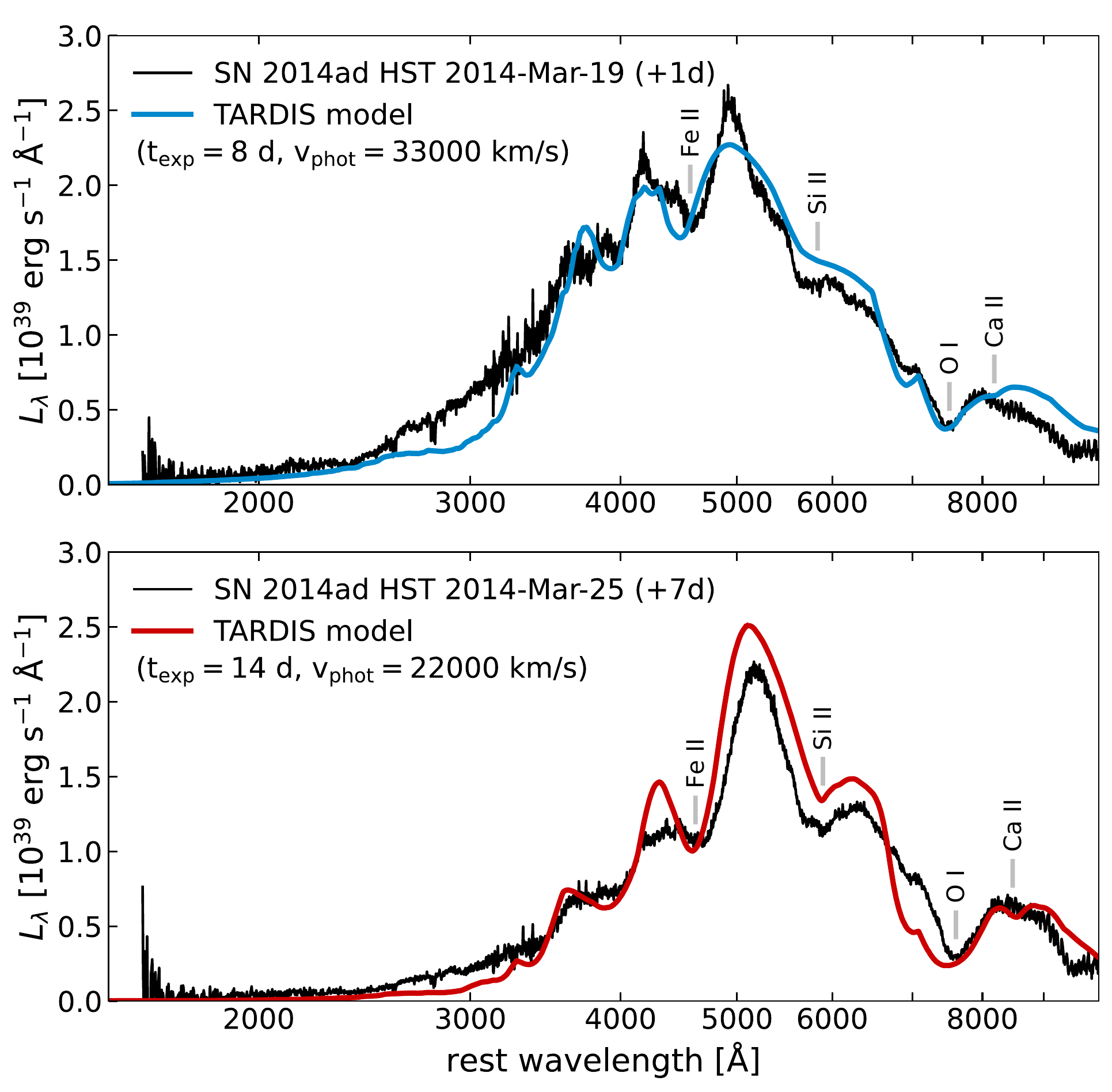}
    \caption{Best-fit \tardis\ model compared to the \textit{HST} spectrum (black) for 2014-Mar-19 (\textit{top}, blue) and 2014-Mar-25 (\textit{bottom}, red). The 2014-Mar-19 model uses $t_{\text{exp}}=8$ d, $v=33,000$ \kms, and $L_{\text{tot}} = \num{5.0e42}$ erg s$^{-1}$. The 2014-Mar-25 model uses $t_{\text{exp}}=14$ d, $v=22,000$ \kms, and $L_{\text{tot}} = \num{4.1e42}$ erg s$^{-1}$. \label{fig:tardis_HST}}
\end{figure*}

\section{Spectral fitting with \tardis \label{sec:models}}

In \autoref{sec:TARDIS} we introduce the radiative transfer code, \tardis, that we use to generate synthetic `model' spectra for SN~2014ad. We detail our \tardis\ settings and parameter choices in \autoref{sec:settings} and present the results of our models in \autoref{sec:results}.

\subsection{\tardis \label{sec:TARDIS}} 

\tardis\ is an open-source, iterative Monte Carlo radiative-transfer code that models SN ejecta in one-dimension \citep{Kerzendorf2014}\footnote{\url{https://github.com/tardis-sn/tardis}}. Running in minutes, \tardis\ is much less computationally expensive than a hydrodynamical simulation with full radiative transfer. Furthermore, \tardis\ produces more physically realistic synthetic models than line-identification codes such as \textsc{synow/syn++} \citep{Fisher2000, Thomas2011}. Like other rapid modelling software, \tardis\ is time-independent and assumes homologous expansion, spherical symmetry, and a sharp blackbody photosphere. Photon packets are injected at the photosphere and their propogation through the model ejecta atmosphere is followed by the code. From user specification of ejecta density structure and composition, \tardis\ calculates self-consistent ionization and excitation states as photon packets interact with ejecta material. Detailed descriptions of \tardis's approach, assumptions, and calculations are presented by \citet{Kerzendorf2014}. 

\tardis's ability to track the photon packet interactions also allows the user to identify which ions and elements contribute to any given spectral feature. This is of particular interest for SN~2014ad where the high velocities cause the spectral lines to broaden and blend together. We create synthetic SN spectra with \tardis\ to better understand the physical properties of SN~2014ad that lead to its spectral evolution.

\subsection{\tardis\ Settings \& Parameters \label{sec:settings}} 

\begin{deluxetable}{cc}[b]
    \tablecaption{SN~2014ad \tardis\ Model Settings \label{tab:tardis_params}}
    \tablehead{
    \colhead{\tardis\ Setting} &
    \colhead{Value}
    }
    \startdata
    Ionization & \texttt{nebular}\\
    Excitation & \texttt{dilute-lte}\\
    Radiative Rate & \texttt{dilute-blackbody}\\
    Line Interaction & \texttt{macroatom}\\
    Number of Iterations & 20\\
    Number of Packets & 100,000\\
    \enddata
\end{deluxetable}

Following the same method as other studies using \tardis\ for spectral fitting \citep[see e.g.,][]{Magee2016,Izzo2019,Barna2021,Williamson2021}, we adopt the parameter settings shown in \autoref{tab:tardis_params} for treatment of the micro-physics in our simulations. 

Given these settings, to generate a synthetic spectrum with \tardis\ we first establish a density profile and elemental abundance profile and then at each desired epoch we input the time after explosion, a requested luminosity, and the photospheric velocity. We adopt a power-law density profile for our \tardis\ model, assuming homologous expansion, with
\begin{equation}
    \rho(v,t) = \rho_0 \left(\frac{v}{v_0}\right)^{-\alpha} \left(\frac{t}{t_0}\right)^{-3} \label{eqn:density}
\end{equation}
where $\rho_0$ is the inner density at the chosen reference time $t_0$, $t$ is the time since explosion, $v$ is the photospheric velocity, $v_0$ is the reference velocity, and $\alpha$ is the power-law index. We choose $\alpha = 6$ based on the success of this profile for the \tardis\ models of SN~2017iuk, a SN-GRB, by \citet{Izzo2019}. For simplicity, we assume uniform fractional element abundances for our \tardis\ models. We find that the major contributing elements to the spectral features are O, Mg, Si, S, Ca, Ti, Cr, Fe, Co, Ni. Despite having little effect on the emergent spectrum, we include C in our model because of its high predicted abundance \citep{Iwamoto1998,Izzo2019}. We specify the initial abundance of $^{56}$Ni directly in the \tardis\ input file and the code appropriately adjusts the relevant abundances to account for radioactive decay since explosion time. \tardis\ does not produce time-evolved models and does not account for the deposition of radioactive energy; rather, \tardis\ simply adjusts abundance levels for a particular time. Other elements have negligible effects on the spectrum so we omit them. We determine initial values for the luminosity and time after explosion from the photometric data \citep{Sahu2018} and estimate the photospheric velocity based on spectral features (e.g., \autoref{fig:FeVels}).

\subsection{\tardis\ fits to SN~2014ad \label{sec:results}} 

We model both epochs of \textit{HST} Optical+UV data (2014 Mar 19, phase $+1$d; 2014 Mar 25, phase $+7$d) with \tardis\ and present our `best-fit' models in \autoref{fig:tardis_HST}. We optimize the models to match the optical wavelengths and compare the predicted model behavior to our data (see Section \ref{sec:tardis_uv}). Because of the complexity of the spectral features and shapes, we evaluate the `goodness-of-fit' qualitatively by eye, rather than using a quantitative technique. We aim to capture the location, widths and relative heights of the major spectral features as well as the overall continuum shape. Some judgements and compromises must be made to determine which parameters best fit the full spectrum, over both epochs. Furthermore, the models have a large number of free parameters (individual element abundances, etc.), some of which can create degenerate effects. So while our models and observed spectra show good agreement, the derived model parameters are not necessarily a unique solution. We discuss this point further in \autoref{sec:discussion}.

\begin{figure}[b]
    \centering
    \includegraphics[width=0.48\textwidth]{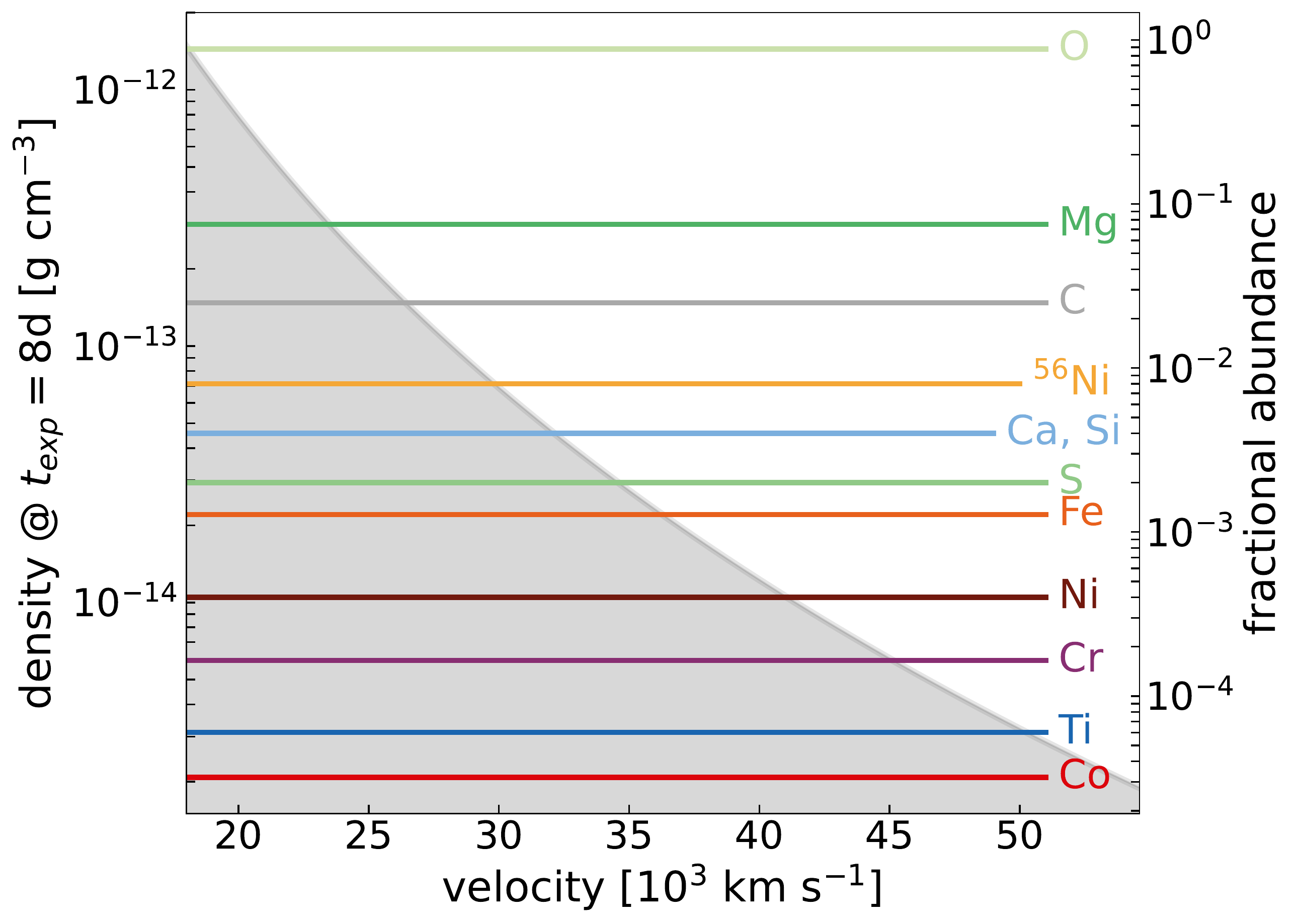}
    \caption{The density profile of the ejecta at $+8$d post-explosion is plotted and filled in gray. A simple power-law, homologous profile is assumed, with $\rho \sim v^{-6}$. The mass fraction abundances of the elements used in our \tardis\ model are overplotted. We assume uniform abundances throughout the SN ejecta. \tardis\ appropriately accounts for the change in Fe-group elements as $^{56}$Ni radioactively decays over time. Element colors match those in \autoref{fig:SDEC_0319}. \label{fig:abundances}}
\end{figure}

Our adopted density and abundance profiles to best match the spectra are shown in \autoref{fig:abundances}. For $t_0 = 8$ days past explosion, we find $\rho_0 = \num{6.9e-14}$ g cm$^{-3}$ at a reference velocity $v_0 =$ 30,000\kms. Importantly, for physical consistency these profiles are used at all epochs. The phase, luminosity, and photospheric velocity model parameters are chosen appropriately for the two epochs. The values used for the two \textit{HST} epochs are given in \autoref{fig:tardis_HST}; the \tardis\ configuration files for these models are publicly available.\footnote{\url{https://github.com/tardis-sn/tardis-setups}}

\autoref{fig:abundances} shows the composition is dominated by O, as expected from the explosion of a massive star with a carbon/oxygen core. Following \cite{Izzo2019}, we group abundances of (Si, S), (Ca, Ti, Cr), and (Fe, Co, Ni \& $^{56}$Ni) in fixed ratios to reflect typical theoretical predictions for hydrodynamical mixing of different burning zones. This helps us to reduce the number of free parameters in our \tardis\ modeling. To get the best fits of particular lines, we alter the ratios of elements within these groups. We did not find a compelling need to deviate from uniform abundances over the velocity range covered by the modeled spectra.

\begin{figure}[h]
    \centering
    \includegraphics[width=0.48\textwidth]{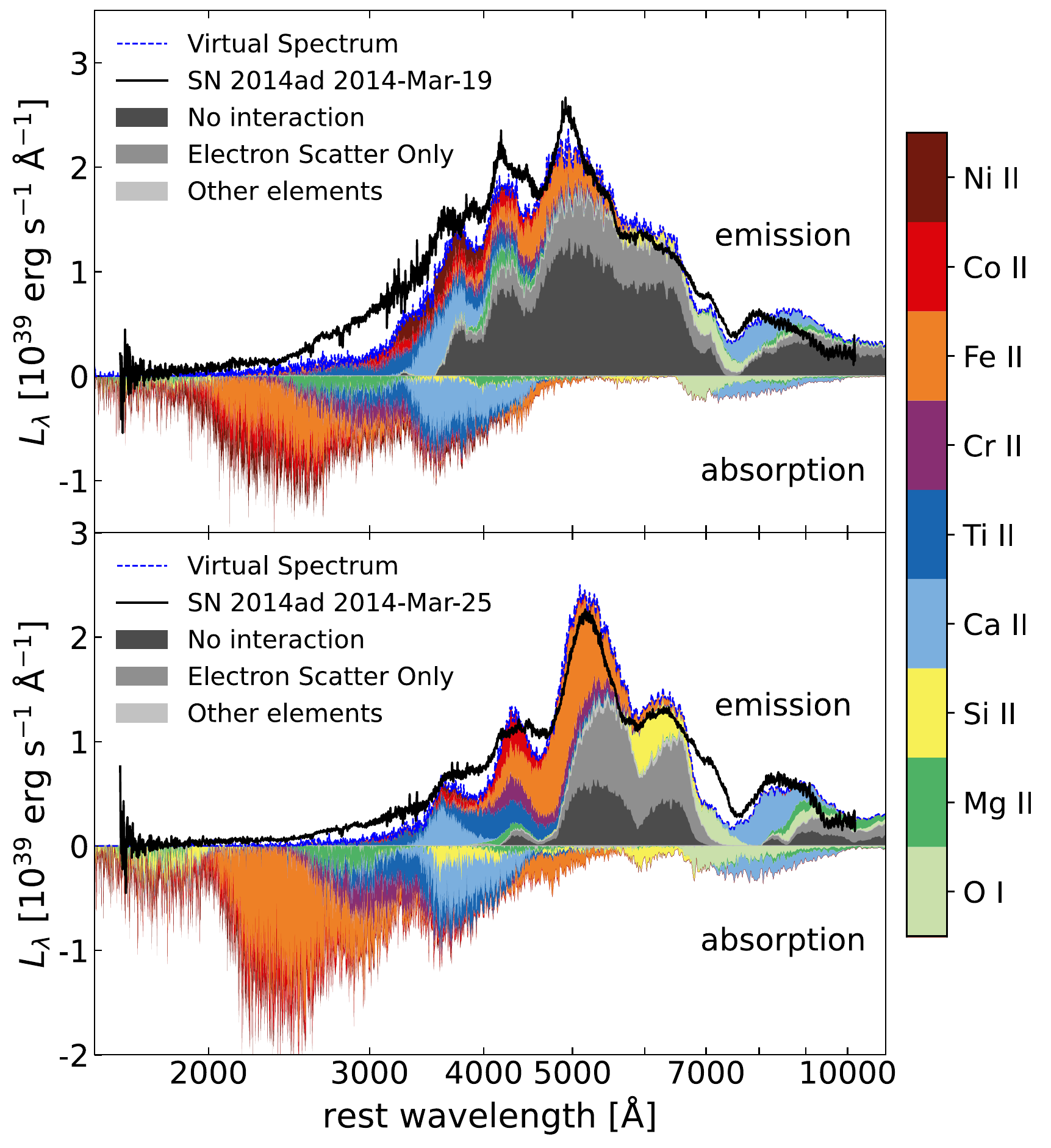}
    \caption{Spectral element decomposition plots for our \tardis\ models of the \textit{HST} spectra. These plots show the contribution of particular elements to spectral features and were generated through a visualization tool available in \tardis. \label{fig:SDEC_0319}}
\end{figure}

The high ejecta velocities in SN~2014ad make it difficult to determine which elements cause particular broad, blended features. However, there are a few key areas controlled predominantly by certain elements or element groups. The largest peak at $\sim$4800 \AA\ in both epochs is dominated primarily by Fe. The dip and shoulder feature around 5700--6100 \AA\ is controlled in shape by Si. The absorption feature at $\sim$7300 \AA\ is a mixture of O I and Ca II. The features blueward of $\sim$4600 \AA\ are contributed to most heavily by Fe, Ni, Co, Cr, Ti, and Ca. The contributions of particular elements to the spectral features are shown visually in \autoref{fig:SDEC_0319}. 

\citet{Sahu2018} showed that inclusion of He marginally improved their \textsc{syn++} fits of SN~2014ad around 5000$-$5600 \AA\ at early times of $+0$d and $+3$d. We explored the addition of He to our \tardis\ models at $+1$d and $+7$d, using the ``recomb-nlte" setting for helium treatment available in \tardis. This setting assumes the He is ionized due to non-thermal excitation, which requires strong Ni mixing in the ejecta; our uniform abundance model inherently satisfies this strong Ni mixing assumption. One caveat is that at early times ($<$2 weeks post-maximum), this modeling assumption may not be correct \citep{Dessart2012}. We find that the addition of even large mass fractions of He does not significantly affect the optical spectra at these early times, and we cannot reproduce the effect from He I in the 5000$-$5600 \AA\ region of the \textsc{syn++} fits by \citet{Sahu2018} with \tardis. However, the radiative temperatures in the ejecta have a significant effect on the plasma state. Our \tardis\ models have a photospheric, or blackbody, temperature of 8040 K at $+1$d and 7915 K at $+7$d; whereas the \textsc{syn++} fits have a blackbody temperature of 13,000 K at $+0$d and 10,000 K at $+3$d. These differences in photospheric temperatures of the models might explain the differing effects from He.

It is difficult to conclusively say anything about the presence of He in the ejecta from our \tardis\ models because He signatures usually strengthen about two weeks after peak \citep{Liu2016}, but our \tardis\ models probe early epochs. Additionally, the line blending from the extremely high velocities prevents us from placing meaningful constraints on the abundance of He. In our modeling with \tardis\, we find that even large amounts of He do not contribute significantly to the early spectra, meaning that the He abundance is unconstrained and large amounts of He could potentially be hidden in the ejecta. Indeed, later time NIR spectra of SN~2014ad show a clean He line in the 2 $\mu$m region \citep{Shahbandeh_2022}; however, constraining the amount of He and explaining why the He features do not appear until later times at lower velocities would require a detailed, multi-zone \tardis\ model to constrain the Ni mixing, which is beyond the scope of this work. The models presented here omit He and do not use the ``recomb-nlte' setting because these He settings do not appear to affect the early spectral model, though we note that significant amounts of He could be hidden and would require modeling of NIR spectra to investigate further.

For our \tardis\ models, the velocities at each epoch (33,000 \kms\ at $+1$d; 22,000 \kms\ at $+7$d) were chosen for the best overall model fit. The model velocities are close to those found by \citet{Stevance2017} and \citet{Sahu2018} for the Si II line velocities at these epochs and within the error of our measured Fe II line velocities (see \autoref{fig:FeVels} and \autoref{tab:FeVels}). In the red ($>$5000 \AA), the model velocities match the natural minima and maxima of the spectra quite well; however, in the blue ($<$5000~\AA) the model minima are somewhat blueshifted, especially in the first epoch. Due to our simplifying assumption of uniform abundances at all velocities, it is difficult to perfectly match all parts of the spectrum. Line-identification codes such as \textsc{synow/syn++} \citep{Fisher2000, Thomas2011} may fit the natural minima and maxima of the spectra better because different velocities can be given for different lines; however, these codes do not enforce a consistent physical structure like \tardis\ does.

One of the challenges in our initial \tardis\ modeling of SN~2014ad was to simultaneously fit the blue and red optical flux. At a fixed time from explosion, the high ejecta velocities imply a large photospheric radius, and for a fixed luminosity, the model effective temperature was cooler than the data suggest, overpredicting the flux redward of 5000 \AA\ and underpredicting the flux blueward compared to the observations. To get a higher model effective temperature, we need a smaller photospheric radius, a higher luminosity, or both. We were able to accommodate this by reducing the time from explosion slightly, putting the first epoch \textit{HST} spectrum (2014 Mar 19) at $t_\text{exp} = 8$ d, smaller than, but consistent at 1$\sigma$ with, the estimate from \citet{Sahu2018} that this epoch should be at $t_\text{exp} = 11 \pm 3$ d. Similarly, our model of the 2014 Mar 25 spectrum adopts $t_\text{exp} = 14$ d to give the best fit. Combined with this shift in explosion date, we also require a higher luminosity for SN~2014ad, adopting a distance of $d = $ 30.1 Mpc for the host galaxy PGC 37625/MRK 1309, an 8\% increase relative to the NED value (28 Mpc) and 14\% higher than the value adopted by \citet{Sahu2018}. Such an increase is plausible: the galaxy is too near to be in the smooth Hubble flow, and thus the redshift-based distance estimate could be significantly uncertain. The preference in the observations for a higher temperature and more blue flux drive the need for a shorter time from explosion and a higher luminosity.

\begin{figure}
    \centering
    \includegraphics[width=.5\textwidth]{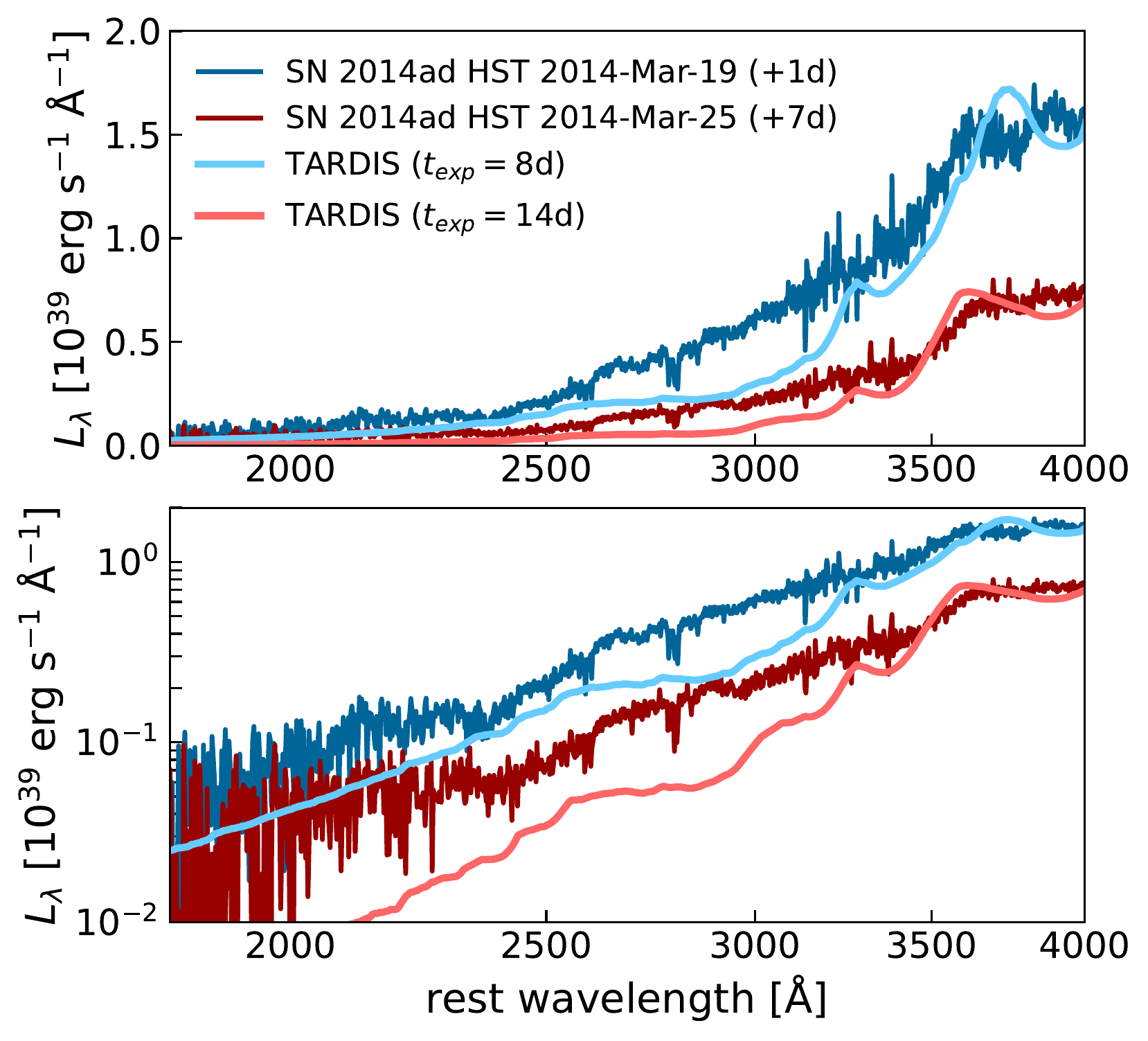}
    \caption{Similar to \autoref{fig:tardis_HST}, but zoomed-in to the UV wavelength range. \textit{HST} data from 2014 Mar 19 is shown in dark blue with its corresponding \tardis\ model in light blue. \textit{HST} data from 2014 Mar 25 is shown in dark red with its corresponding \tardis\ model in light red. The bottom panel is the same as the top plot, but with a log scale in luminosity. The \tardis\ models are featureless and under-predict the observed flux in the UV. \label{fig:tardis_uv_compare}}
\end{figure}

\subsubsection{UV Comparison \label{sec:tardis_uv}}
We deliberately did not tune the \tardis\ models to fit the UV data, preferring instead to estimate the best fit from the optical only and then compare the model prediction in the UV to the novel \textit{HST} observations. \autoref{fig:tardis_uv_compare} highlights the model comparison to the data in the UV. Even though we pushed the model parameters to yield more blue flux relative to red in the optical, the models still significantly under-predict the flux in the UV wavelengths. This is especially the case around 2600--3100 \AA, where there appears to be a broad flux deficit in the model that is not seen in the data. Neither the model nor the data show distinct spectral features in this region: \autoref{fig:SDEC_0319} shows strong UV line-blanketing from Fe-group elements, Ca, Cr, and Ti. The lack of features and extreme line blending makes it challenging to improve the model's UV fit. Complicating things further, the same elements that contribute in the UV have major impacts on the optical spectrum. Thus, we find that, assuming uniform abundances, improving the UV fit can compromise the optical fit. 

Because of the higher optical depth from the forest of lines, the UV wavelengths are very sensitive to metallicity and probe the higher velocity, outer layers of the ejecta, so future improvements to the model in the UV might be achieved by more sophisticated modeling with non-uniform abundances at higher velocities \citep[see e.g. ][]{Barna2021}. Additionally, the UV might be better fit with a more complicated density profile with a different high-velocity power-law slope. Exploring such possibilities is beyond the scope of the current work, as these changes are too unconstrained. Future theoretical models may provide a clear path forward, but we currently have no independent method to check whether any modifications we make to the model in the UV would be a correct or unique solution. \citet{Ben-Ami2015} suggest that the UV excess in most of their sample of SN IIb may come from interaction with a circumstellar medium (CSM). However, the difference between our observed and model spectra in the UV for SN~2014ad is much smaller than for the SNe IIb in \citet{Ben-Ami2015} so we do not find a compelling need for an additional CSM component for SN~2014ad. Other contributing factors to the discrepancy between the model and the UV data may be that the \tardis\ models are subject to larger amounts of Monte-Carlo noise in this low-flux region and that uncertainties in the extinction can substantially affect the UV.

We do not extend the \tardis\ model to compare to the NIR spectra because the assumption of a sharp blackbody photosphere, employed by \tardis, breaks down in the IR due to low opacities.

\subsubsection{Multi-Epoch \tardis\ Models \label{sec:multi-epoch}}

\begin{figure}
    \centering
    \includegraphics[width=0.48\textwidth]{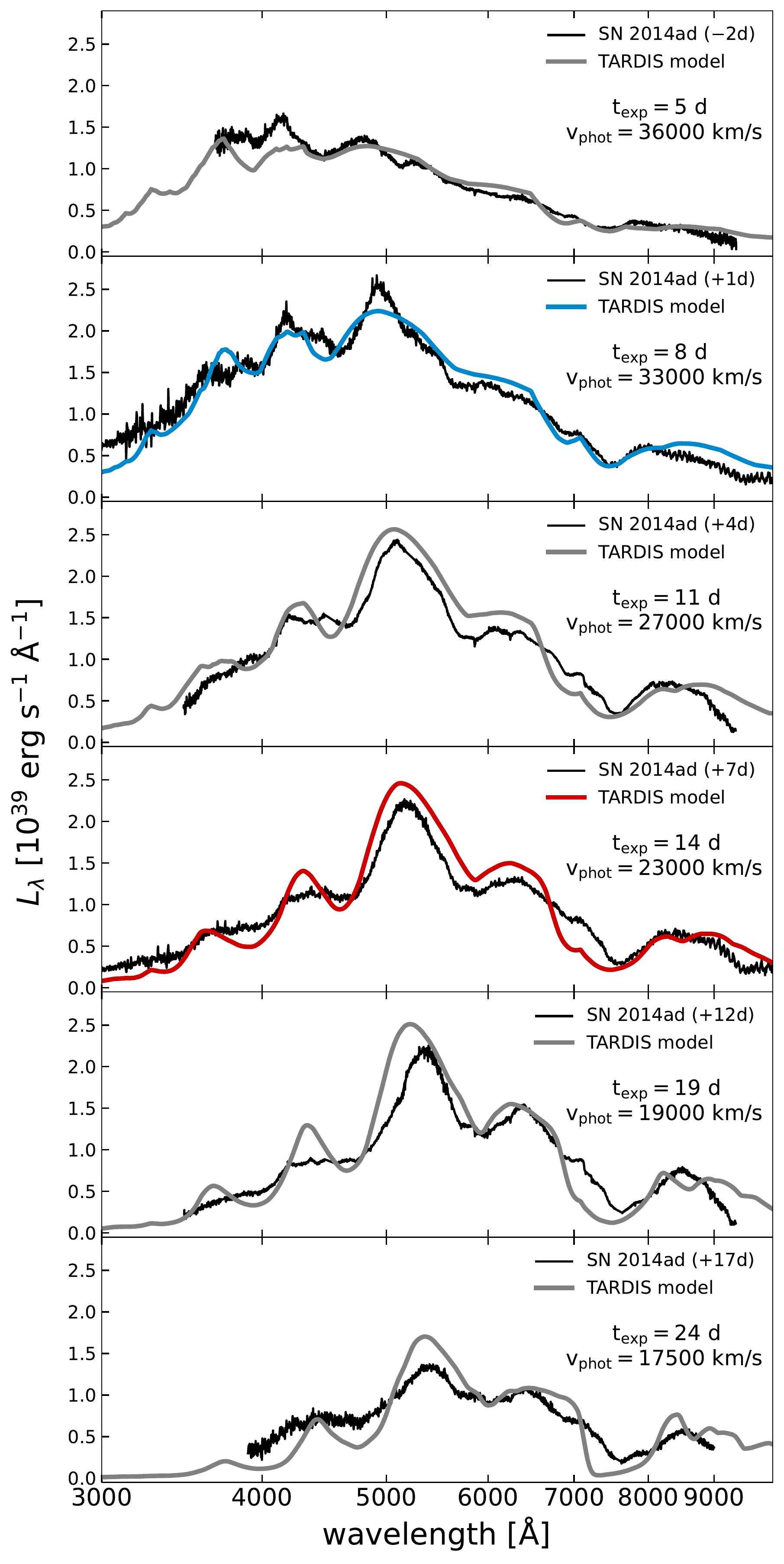}
    \caption{\tardis\ models were optimized to match our \textit{HST} Optical+UV data epochs (blue and red; same as in \autoref{fig:tardis_HST}). Here we show the model prediction compared to the data at other epochs, with the photospheric velocity as the only free parameter for each additional panel. \label{fig:multi_epoch}}
\end{figure}

Our \tardis\ model was optimized to match the optical spectra at the two \textit{HST} epochs. We can extend this by comparing the model prediction to additional epochs with observed optical spectra. We use the same element abundances and density profile and maintain the explosion date shift and luminosity scaling used in the \textit{HST} Optical+UV \tardis\ models discussed in Section \ref{sec:results}. Thus, the photospheric velocity is treated as the only free parameter in these additional model epochs. \autoref{fig:multi_epoch} shows the results of extending our model to one earlier, one in-between, and two later epochs. The models match the observed spectra quite well, accounting for the major features and relative strengths. If we go beyond the epochs shown in \autoref{fig:multi_epoch}, the model begins to deviate substantially from the observed spectra. Fitting these very early and late times may require departure from our simplifying assumptions of uniform element abundances and/or a single power-law density profile with $\alpha = 6$. Such an investigation is beyond the scope of this paper and will be the subject of future work.

\subsubsection{Density and Mass Constraints \label{sec:tardis_mass}}

We can use the density profile derived in our \tardis\ model to place constraints on the SN ejecta mass. We estimate mass by integrating the \tardis\ density profile over a velocity range (for total mass, $v_{\mathrm{min}} \sim$ 1000 \kms\ and $v_{\mathrm{max}} \sim$ 100,000 \kms). To constrain the mass estimate we vary the density profile normalization parameter $\rho_0$ to determine loose lower and upper bounds beyond which the models give poor fits to the data. This is a qualitative exercise, so we are generous in our evaluation of these bounds, especially because density is somewhat degenerate with photospheric velocity and element abundances (particularly those that contribute heavily at $<5000$ \AA) in the model spectra. Nevertheless, there are limits on this degeneracy and certain $\rho_0$ values will not yield a reasonable fit to both epochs, regardless of other parameter changes.

Our \tardis\ models for the two \textit{HST} epochs probe the velocity region of 22,000--50,000\kms. We loosely estimate a lower limit of $\rho_{0\textrm{,lower}} = 0.3 \rho_{0\textrm{,best}}$, and an upper limit of $\rho_{0\textrm{,upper}} = 3 \rho_{0\textrm{,best}}$. The corresponding lower, best, and upper limit SN mass estimates from integrating the density profile (assuming spherical symmetry) in this velocity region directly probed by the data are 1.0 M$_\odot$, 3.0 M$_\odot$, and 9.0 M$_\odot$, respectively.

Our two epochs of modeled data do not probe the density profile at velocities below 22,000\kms\ or above 50,000\kms. While the density at higher velocity does not contribute a substantial amount of additional mass, a continued sharply rising density profile with $\alpha = 6$ to lower velocity would imply a huge reservoir of interior mass. From physically realistic expectations for the total mass and density models based on simulations such as the CO138E30 model from \cite{Iwamoto1998}, we expect the density profile to flatten at low velocities. The CO138E30 model has a gradual flattening beginning around 15,000\kms\ reaching approximately constant density by around 10,000\kms. Here we model this behavior as a sharp change from a power-law to a constant profile for simplicity, even though realistically, we would expect a smoother transition. Motivated by the turnover in the CO138E30 model, we assume a sharp density flattening at velocity $<$ 15,000\kms\ and calculate a total mass of 20.5 M$_\odot$. If we assume the extreme density profile flattening at $<$ 22,000\kms, just outside the velocity range probed by our models, the total mass becomes 6.5 M$_\odot$. Clearly, the total mass estimates change dramatically depending on this choice of flattening velocity. In all cases they still imply that there is a large amount of additional mass at low velocities.

In our density profiles, we choose a reference velocity $v_0 =$ 30,000\kms\, within the range probed by our models, so that $\rho_0$ is the driving parameter in the model spectra; varying the power law index $\alpha = 6$ does not have a strong effect at velocities near $v_0$. Indeed, we find that moderate changes to the abundances can produce equally good \tardis\ models with a range of $\alpha$ from $\sim$ 4 to $\sim$ 8. When extrapolated to low velocity, the steeper density slope ($\alpha = 8$) implies even higher mass. The shallower density slope ($\alpha = 4$) results in lower total mass estimates of 9.1 M$_\odot$ and 5.9 M$_\odot$ when flattened at velocities of $< $15,000\kms and $<$ 22,000\kms, respectively. Overall, these results suggest a high ejecta mass and thus a high-mass progenitor for SN~2014ad.

\begin{figure}
    \centering
    \includegraphics[width=0.48\textwidth]{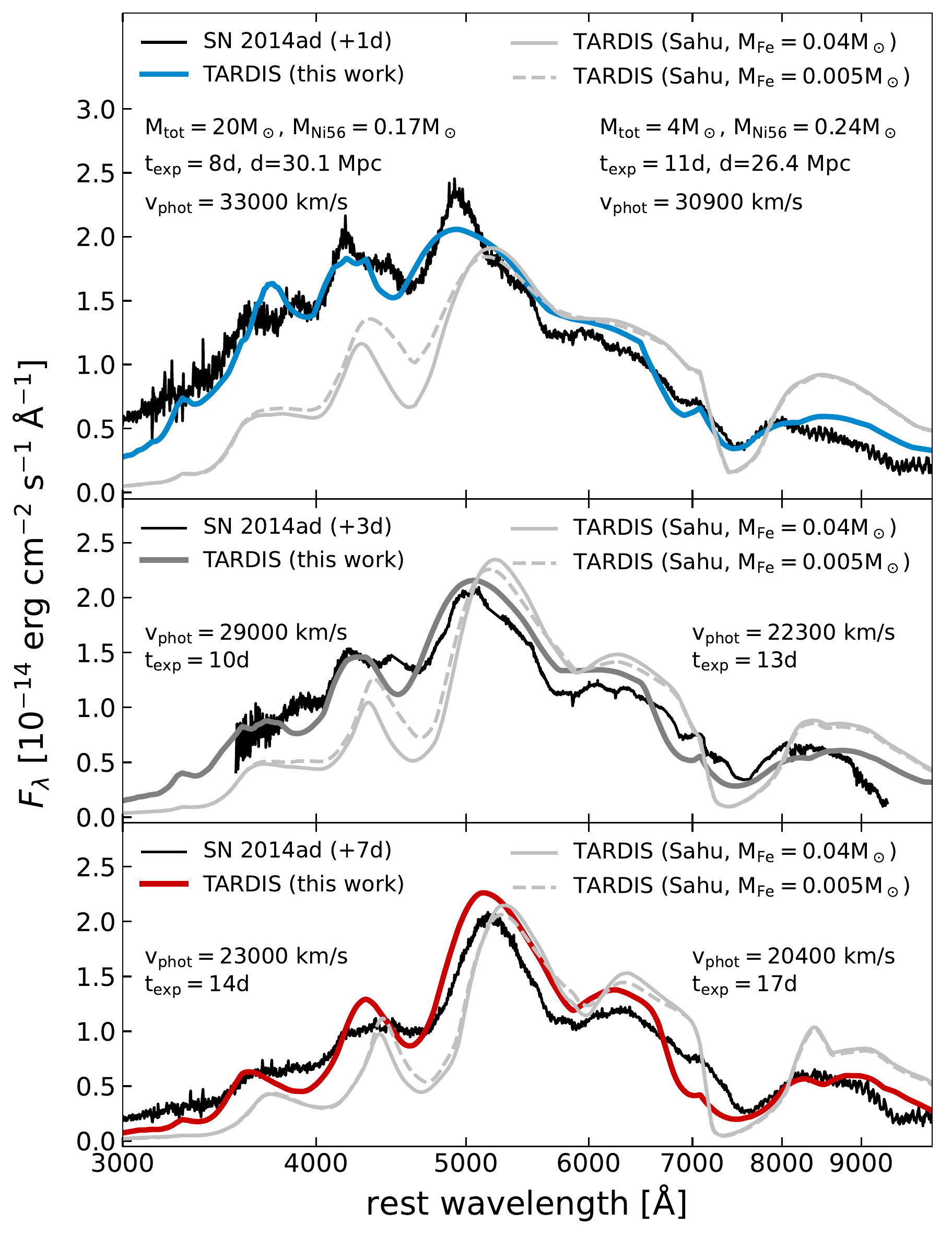}
    \caption{Comparison of our \tardis\ models to the \tardis\ models generated using parameters from the light-curve analysis in \citet{Sahu2018} (light gray). The light gray \tardis\ model has lower density (to match the lower ejecta mass) and higher Fe-group element abundances (to match the higher $^{56}$Ni mass). The dashed light gray model is the same as the light gray model, but only the $^{56}$Ni abundance has been increased rather than all the Fe-group elements. In the dashed light gray model, the Fe abundance is the same as our \tardis\ models. The parameters used in our \tardis\ models are given on the left and the parameters used in the \tardis\ models based on \citet{Sahu2018} are given on the right. \label{fig:sahu_models}}
\end{figure}

Surprisingly, the high density that we find necessary in our \tardis\ models results in high mass and kinetic energy estimates that are at odds with those derived from the light-curve analysis by \citet{Sahu2018}. They find $M_\text{ej} = 3.3 \pm 0.8 \; M_\odot$, $E_k = (1.0 \pm 0.3) \times 10^{52}$ erg, and $M_\text{Ni56} \simeq 0.24 \pm 0.07 \; M_\odot$. In contrast, for our best-fit density model and assuming a flattening at 15,000\kms\ (motivated by the CO138E30 model), we derive $M_\text{ej} \simeq 20.5 \; M_\odot$, $E_k \simeq \num{8e52}$ erg, and $M_\text{Ni56} \simeq 0.17 \; M_\odot$, a significantly higher ejecta mass and kinetic energy (by a factor of several), yet a slightly lower $^{56}$Ni mass. Even our lowest mass estimate, with $\alpha = 4$ and a flattening at 22,000~\kms, gives $M_\text{ej} \simeq 5.9 \; M_\odot$, $E_k \simeq \num{9.5e52}$ erg. To investigate this discrepancy, we run \tardis\ models directly using the light-curve based parameters from \citet{Sahu2018}. We scale our density profile and Fe-group element abundances such that the total mass and $M_\text{Ni56}$ are consistent with the values from \citet{Sahu2018}.

\autoref{fig:sahu_models} shows the spectra predicted by \tardis\ for the parameters from \citet{Sahu2018} in light gray. While these spectra produce roughly the correct features and shapes, the model significantly under-predicts the observed flux at bluer wavelengths ($<5000$ \AA), and the Ca II features around 7500--8500 \AA\ are not captured as well. To illustrate that the low model flux in the blue wavelengths is not just due to increasing the Fe abundance, we show the \tardis\ spectra produced by only scaling up the $^{56}$Ni abundance, and not also the other Fe-group elements, in dashed light gray.

Our models match the observed spectra more closely than the \tardis\ models based on the parameters derived from the light-curve analysis by \citet{Sahu2018}, but they predict a higher ejecta mass than has been typically estimated for the class of SN Ic-bl. Moreover, since SN Ic-bl have been stripped of their hydrogen and helium, the supernova ejecta reflects the progenitor's carbon/oxygen (CO) core mass. The mass derived from our best fit with a velocity flattening around 15,000~\kms, $\sim$20.5 M$_\odot$, exceeds the largest CO core mass ($\sim$ 14 M$_\odot$) in the stellar evolutionary models calculated by \citet{Sukhbold2016}, achieved for a progenitor mass, M$_{\text{ZAMS}} \sim$ 40 M$_\odot$ at solar metallicity \citep{Sukhbold2016, Katsuda2018}. This theoretical CO core mass limit can be increased much higher ($\lesssim$ 40 M$_\odot$) for the case of much lower metallicity \citep{Heger2003} or the chemically homogeneous model \citep{Yoon2005}, which results from rapid rotation and low metallicity and is a popular scenario for long GRB progenitors. However, the metallicity of SN~2014ad ($\sim$ 0.5 Z$_\odot$) is not as low as required for a very high CO core mass, and our models likely cannot be reconciled with the theoretical CO core mass limit in this way. Thus, an ejecta mass of $\sim 20.5$ M$_\odot$ would require several solar masses of leftover hydrogen and/or helium on top of the CO core. 

The ejecta mass directly probed by our modeled epochs is limited to the velocity range 22,000 $< v <$ 50,000\kms; within that range the mass estimate is 3.0 M$_\odot$, comfortably below the maximum CO core mass. Thus, one way to bring our best-fit \tardis\ model into agreement with the theoretical CO mass limit is to require a shallower $\alpha$ or flatten the density profile beginning at higher velocity. As described above, these modifications can lead to lower mass estimates $\sim$ 6 M$_\odot$, which are consistent with theoretical CO mass limits. The lowest conceivable mass we can derive from our \tardis\ model density profile is $\sim$ 4.5 M$_\odot$, which would require $\alpha = 6$ and flattening at $v \lesssim$ 25,000\kms, which enters the velocity region probed by our models. Modifying the density profile in this way has minor effects on our \tardis\ model. The first epoch is unaffected because the velocities it probes are $>$25,000\kms; the second epoch is slightly affected, but discrepancies could be remedied with changes to the abundances. While these modifications can resolve tensions between the mass predicted by our models and theoretical CO mass limits, they still produce masses a factor of two larger than the ejecta mass calculated by \citet{Sahu2018} from light-curve data.

We attempt reconciliation with the values derived from the photometry by considering that the light curve directly constrains the diffusion time-scale $\tau_m$ \citep{Arnett1982} given by 
\begin{equation}
    \tau_m = \bigg(\frac{\kappa}{\beta c}\bigg)^{1/2} \Bigg(\frac{6 M_{\mathrm{ej}}^3}{5 E_{\mathrm{k}}}\Bigg)^{1/4}
\end{equation}
where $\kappa$ is the optical opacity, $\beta$ is a constant of integration, and $c$ is the speed of light. Thus, the light-curve analysis gives a constraint on the ratio of $M_{\mathrm{ej}}^3/E_{\mathrm{k}}$, which is proportional to $\tau_m$. We calculate the velocity at which our density profile would need to flatten in order to match the value of $M_{\mathrm{ej}}^3/E_{\mathrm{k}}$ from \citet{Sahu2018}. For $\alpha = 6$ (used in our best-fit models), we calculate a required flattening at 23,000\kms, corresponding to a total mass of 5.6 M$_\odot$. We repeat this exercise, calculating the flattening velocity required to give the same $M_{\mathrm{ej}}^3/E_{\mathrm{k}}$ ratio derived from the light-curve data, for $\alpha$ values ranging from $3$ (below which the mass will diverge at \textit{high} velocity unless we introduce another cut-off) to $9$. We find that at as $\alpha$ decreases and the density profile becomes less steep, the flattening velocity required to match the $M_{\mathrm{ej}}^3/E_{\mathrm{k}}$ ratio actually decreases, resulting in an increase in total mass due to the large amount of mass at low velocity. For example, in the case of a shallower density slope of $\alpha = 4$, we calculate a flattening velocity of 15,800\kms, which corresponds to M$_{\mathrm{tot}}=$9.2 M$_\odot$. Thus, the observed $M_{\mathrm{ej}}^3/E_{\mathrm{k}}$ ratio can be matched by a wide range of $\alpha$ and flattening velocity values; however, lower mass requires a higher flattening velocity and steeper density slope (e.g. $\sim 4.5$ M$_\odot$, 25,000\kms, $\alpha =$ 8.5). Thus, our models, theoretical CO mass limits, and the light curve analysis can be reconciled if we seek to match $M_{\mathrm{ej}}^3/E_{\mathrm{k}}$ rather than $M_{\mathrm{ej}}$ from \citet{Sahu2018} and we impose a flattening of the density profile. 

If steeper or shallower $\alpha$ and/or flattening the density profile at higher velocity is the right solution to bring the light-curve and spectroscopic data mass estimates into accord, they should have major impacts on later time spectra that directly probe this region. Modeling the late time epochs would be a great test of the density profile; however, at late times, the element abundances require changes as well and quickly become degenerate with the density. Exploring this degeneracy is difficult to do manually and will be the subject of future work. In this work, our best constraint on the density profile at low velocities comes from the nebular line profiles at late times (see Section \ref{sec:nebular_analysis}). We discuss the implications from analysis of the density profile at early times using our \tardis\ models and at late times using the nebular line profiles in Section \ref{sec:discussion}.

Given our simplistic density and abundance profiles, and the limitations of our 1-d model, we cannot rule out the possibility there are other ways to reconcile the high density suggested by the spectra with the total mass estimate derived from the light curve. In our exploration, however, we were not able to find parameters that would generate good spectral fits at lower densities, specifically to the observed blue flux at wavelengths $<$ 5000 \AA.

\section{Nebular Line Fitting \label{sec:nebular_analysis}} 

\begin{figure}[b]
    \centering
    \includegraphics[width=0.48\textwidth]{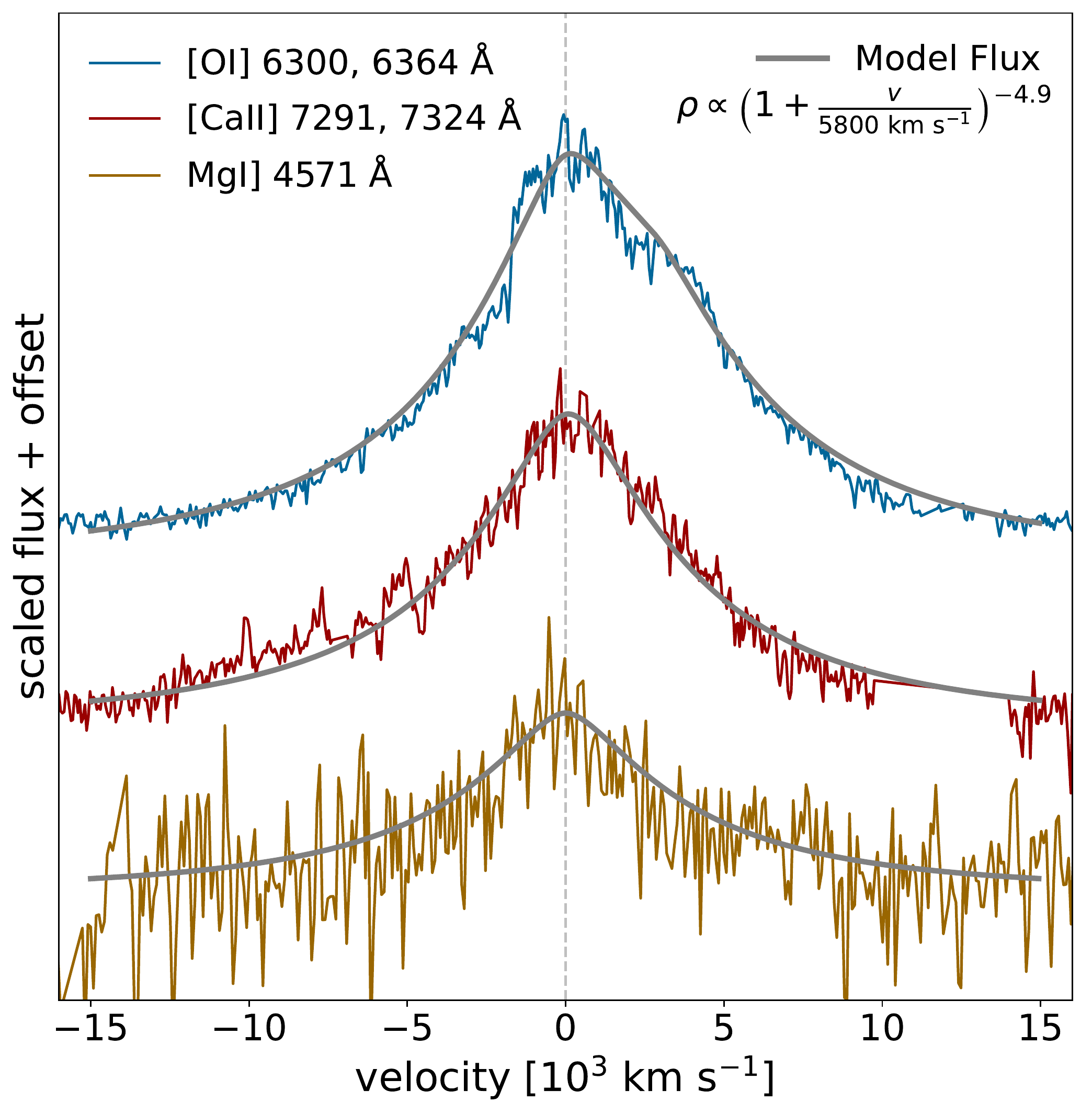}
    \caption{Velocity line profiles of [O I]~$\lambda\lambda$6300,~6364, [Ca II]~$\lambda\lambda$7291,~7324, and Mg I]~$\lambda$4571~\AA. The dashed line marks zero velocity with respect to 6300, 7291, and 4571 \AA, respectively. Narrow host galaxy emission lines and telluric lines have been removed. The modeled flux from integrating over a sphere with density $\rho \propto \left(1+v/v_\mathrm{scale}\right)^{-\alpha}$ and assuming uniform excitation is fit to the nebular line features using MCMC techniques. The [O I] fit gives $v_\mathrm{scale} = $ 6300$^{+2300}_{-1600}$~\kms\ and $\alpha = $ 4.9$^{+0.6}_{-0.4}$; the [Ca II] fit gives $v_\mathrm{scale} = $ 5300$^{+1300}_{-900}$~\kms\ and $\alpha = $ 4.9$^{+1.0}_{-0.7}$; and the Mg I] peak gives $v_\mathrm{scale} = $ 6000$^{+3000}_{-3000}$~\kms\ and $\alpha = $ 6.3$^{+1.8}_{-2.0}$. We overplot the modeled flux in gray, choosing one set of parameters that are consistent with the fits for all three lines.}
    \label{fig:neb_line_vels}
\end{figure}

To investigate the density structure of SN~2014ad at low velocity, we analyze the late time nebular line profiles. In the nebular phase, the opacity of the ejecta drops such that the photons from all regions are free-streaming and the low-velocity, interior layers of the SN are revealed. The emissivity structure of the ejecta, which is a combination of the density and excitation, determines the shape of the nebular line features \citep[for a review, see][]{Jerkstrand2017}. For example, nebular emission from constant density, uniformly emitting ejecta results in features with a parabolic shape; ejecta distributed in a shell produces flat-topped boxy profiles; and ejecta with a torus geometry give rise to double peaked features \citep[e.g., see][]{Modjaz08,Milisavljevic2010}. SN~2014ad exhibits peaky, triangular shaped features, indicating that there must be an increase in emission towards a compact core. This may come from an ejecta density profile that increases towards the center, an extra emission source, or a combination. 

To investigate the density structure of SN~2014ad, we assume spherical symmetry and uniform excitation. Following the calculations described by \citet{Jerkstrand2017} in Section 2.1, we integrate our density profile over a sphere to obtain an expression for the flux. We choose a softened power-law density distribution and fit our expression for the resulting flux to the observed velocity line profiles. Our density profile is given by
\begin{equation}
    \rho(v) \propto \left(1+v/v_\mathrm{scale} \right)^{-\alpha}
\end{equation}
where $v$ is the velocity, $v_\mathrm{scale}$ is a scale velocity, and $\alpha$ gives the power-law slope at high velocity. We adopt this form because it acts like the simple power law profile from our \tardis\ models at large velocities, $v \gg v_\mathrm{scale}$, but it does not diverge at low velocities.

We use the MCMC package {\tt emcee} \citep{emcee} to fit to the double [O I] peak and find that the scale velocity and the power-law slope are tightly correlated. Best-fit values for the parameters are $v_\mathrm{scale} = $ 6300$^{+2300}_{-1600}$~\kms\ and $\alpha = $ 4.9$^{+0.6}_{-0.4}$. Fitting the double [Ca II] peak, we find $v_\mathrm{scale} = $ 5300$^{+1300}_{-900}$~\kms\ and $\alpha = $ 4.9$^{+1.0}_{-0.7}$. These values are in agreement with those derived from the [O I] peak. Additionally, we fit the Mg I] peak and find parameters that agree with those from [O I] and [Ca II], $v_\mathrm{scale} = $ 6000$^{+3000}_{-3000}$~\kms\ and $\alpha = $ 6.3$^{+1.8}_{-2.0}$, though with much larger errors due to the weak line strength and larger noise in the data. Through the MCMC fitting process, the errors in the nuisance parameters (e.g. amplitude and vertical offset) are also accounted for.

Under the uniform excitation assumption, these findings indicate that the density profile is steep at low velocity near the core, implying a large amount of mass in this region. Alternatively, if we assume that the density profile must flatten at low velocities, then our fitting suggests that excitation must increase towards the core. This could be due to an additional energy source in the core on top of the radioactive decay. In the spectra of SN~2014ad, the Mg I]~4571~\AA \ line has a narrow peak which cannot be from the H II region emission, and may support extra emission on the core.  In the context of a SN Ic-bl, this extra emission could potentially be a signature of continuing magnetar energy. However, if the peaky features are from an excitation effect, we might expect different results for the fits of [O I], [Ca II] and Mg I], whereas our fits for each element are consistent with each other. We discuss these results together with our results from our \tardis\ models in Section \ref{sec:discussion}.

\section{Discussion \label{sec:discussion}} 

\begin{figure*}
    \centering
    \includegraphics[width=\textwidth]{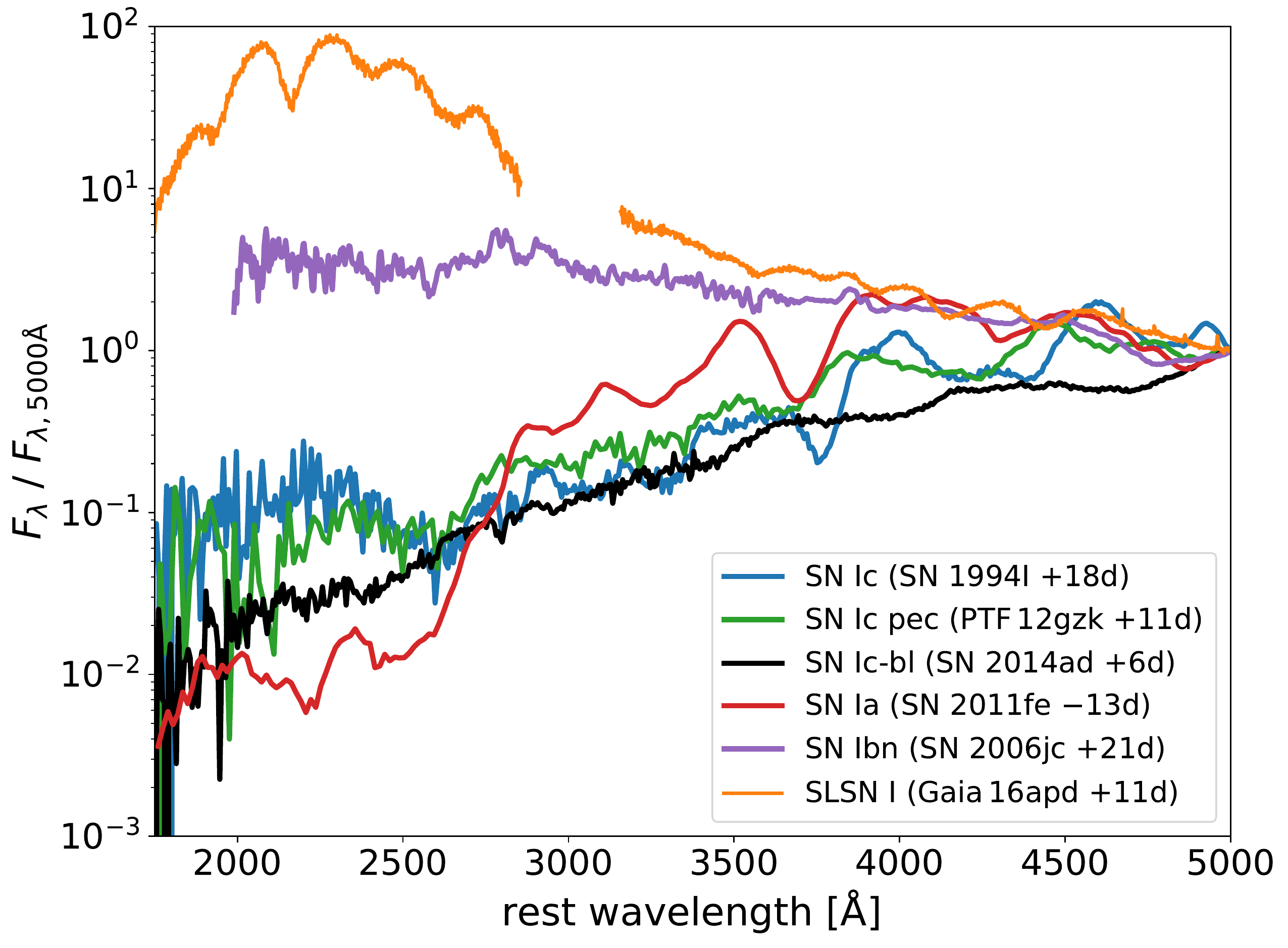}
    \caption{SN 2014ad UV spectra placed in context with other UV spectra of type I supernovae. Spectra have been normalized at 5000 \AA\ to emphasize the relative brightness of the UV to optical flux. The SNe Ic and SN Ic-bl spectra most closely resemble the SN Ia spectrum in the UV. The spectra have been logarithmically rebinned for clarity in the UV. \label{fig:sn_uv_diversity}}
\end{figure*}

Our \textit{HST} UV spectra of SN~2014ad show suppressed flux levels and are nearly-featureless. We interpret this as the effect of high line-blanketing opacity strongly absorbing UV radiation, particularly from Fe-group elements; these elements have many transitions at UV wavelengths, creating a forest of absorption lines. The lack of easily identifiable features in the UV spectra arises from extremely high ejecta velocities in SN~2014ad, broadening and blending the UV absorption lines. This implies the presence of high-density material including Fe-group elements at high velocities, or correspondingly, in the outer layers of the ejecta.

In our analysis, we compare a UV spectrum of SN~2014ad to a UV spectrum of SN~1994I, the ``canonical'' SN Ic, and PTF\,12gzk, a peculiar SN Ic with high blue-shifted line velocities (but without similarly broadened lines). SN~1994I and PTF\,12gzk’s UV spectra look similar to that of SN~2014ad; they share the overall general shape and subdued flux. However, they do exhibit a few more distinct features because SN~1994I and PTF\,12gzk’s lower velocities do not broaden and blend the absorption lines as much. In \autoref{fig:convolution} we show that these spectral differences in the UV largely blend out as the SN Ic spectra are convolved to higher velocities; this suggests a smooth spectral transition from SN Ic to SN Ic-bl in the UV, though we caution with strong caveats given the small sample of just three SN Ic with available UV spectra. A larger UV spectroscopic sample of SN Ic and Ic-bl is needed to explore the diversity of UV behavior for these objects. Interesting UV behavior might also be found at very early pre-maximum time where the UV is strongest. If little UV spectral variety is found for these objects, as we see in this work, it could indicate that (less expensive) photometric observations are sufficient to characterize SN~Ic at UV wavelengths.

Our measurements of the Fe II 5169 \AA\ velocities show that SN~2014ad has unusually high early velocities for a Ic-bl; this is consistent with the analysis in \citet{Sahu2018} derived from the Si II 6355 \AA\ line. We find that SN~2014ad’s velocities are systematically higher than both the sample of SNe Ic-bl without observed GRBs and SN-GRBs. From these results, SN~2014ad appears to fit more consistently with the higher velocities of the SN-GRB sample and we might favor the on-axis-unobserved GRB scenario. However, an accompanying on-axis GRB for SN~2014ad is ruled out by strong X-ray and radio constraints \citep{Marongiu2019}. Nevertheless, it is still possible that SN~2014ad may have harbored an off-axis jet, which is not excluded by the radio limits, especially given the low metallicity environment that SN~2014ad inhabits, similar to the PTF SNe Ic-bl sample of \citet{Modjaz20}. Future work on modeling the passage of off-axis vs. choked jets in envelopes of exploding massive stars coupled with radiative transfer calculations are needed to explain the high ejecta velocities in SN~2014ad given the lack of an associated on-axis GRB.

We model our two epochs of \textit{HST} UV spectra of SN~2014ad with \tardis. Our models assume simple element abundance and density profiles because the spectra do not require more complexity to achieve reasonable fits over an extended timeframe. Our best-fit \tardis\ synthetic spectra require both decreasing the time from explosion to the lower limit allowed by the photometric observations \citep[7 days before \textit{B}-maximum;][]{Sahu2018}, and increasing the luminosity by adopting a 15\% larger distance. Though somewhat uncomfortably in tension with previous results, we find these modifications are necessary to match the observed high velocities and relatively high blue flux at $<5000$ \AA. Indeed, \citet{Sahu2018} show that SN 2014ad’s early spectra are quite blue in comparison to other SN Ic-bl and SN-GRBs. We find that capturing these blue features with our model also requires high densities at high velocities.

We consider the consequences predicted by the high densities preferred by our \tardis\ models. Integration of the density profile yields an ejecta mass of $\sim$20.5 M$_\odot$ (assuming a density flattening for velocities $<$ 15,000\kms, a turnover chosen based on the CO138E30 model from \citet{Iwamoto1998}). Such a large mass is inconsistent with both the light-curve analysis ejecta mass estimate and the largest theoretically derived CO core mass from \citet{Sukhbold2016} (assuming solar metallicity). Reconciliation of the mass estimate with the $M_{\mathrm{ej}}$ from the light curve analysis can be achieved if we impose a flattening in the density profile at higher velocity. Alternatively, if we seek instead to reconcile with the observed $M_{\mathrm{ej}}^3/E_{\mathrm{k}}$ ratio from the light curve diffusion timescale, a wider range of power-law slopes and flattening velocities is allowed. In this case, steeper slopes require higher flattening velocities and yield lower total mass estimates, while shallower slopes require lower flattening velocities and yield higher total mass estimates. Modification of the density profile for $4 \lesssim \alpha \lesssim 8$ and flattening velocity $<$ 25,000\kms\ does not impact our \tardis\ models significantly and presents a possible solution to the inconsistency. However, the choice of $\alpha$ and flattening velocity (especially at high velocities) should substantially alter the model fits at later epochs which probe lower velocity regions of the ejecta. Further modelling of the late time spectra would help constrain the density profile at lower velocities, and will be the subject of future work.

Extrapolating the density profile to velocities below those probed by our models is very uncertain. However, regardless of what happens at low velocities, our \tardis\ model directly predicts $\sim$ 3 M$_\odot$ of ejecta at high velocity in the range 22,000\kms\ to 50,000\kms. This is the same amount of total ejecta mass estimated from the light-curve analysis and it has intriguing implications for the explosion model for SN~2014ad. The mass at high velocity ($v >$ 22,000 \kms, probed by the spectra) would be a very large fraction of the total ejecta mass. Thus the engine of the explosion would somehow need to effectively couple to the ejecta, accelerating the bulk of it to quite high velocities, with relatively little material at low velocity.

To investigate the density profile at low velocity, we place the results of our nebular line fitting in the context of our \tardis\ models. Assuming uniform emission from the ejecta, we find that the shape of the nebular features prefer a density profile with a slope of $\alpha \sim$5 and a scaling velocity $\sim$ 5000~\kms. While we caution that this scaling velocity is not the same as the flattening velocity suggested in Section \ref{sec:tardis_mass}, these parameter values suggest that the density profile is still steep at low velocities and there is indeed significant mass at low velocity. Furthermore, the peaky nebular features disagree with our suggested flattening and also with the more realistic smooth turnover from the CO138E30 model \citep{Iwamoto1998}. A nearly constant, or shallowly sloped density profile at low velocity would produce nebular features that are more parabolic than we observe. Interestingly, the fit to the nebular lines yields a value for $\alpha$ that is in the range preferred by our \tardis\ models. Because of this, if we scale the best-fit nebular density profile to the $\rho_0$ used in our \tardis\ models and input it into \tardis, we recover good \tardis\ models. However, this density profile from the nebular fits, scaled to $\rho_0$, would predict an unphysical total mass of $\sim$ 19 M$_\odot$.

To solve a similar problem for Ic-bl SN~1998bw, \citet{Maeda2006} invoked a ``two-component" density profile where the roughly power-law density at high velocities flattens off around 15,000~\kms, and then steepens again at nebular velocities before flattening for a final time around 3,000~\kms. Such a density profile dramatically improved their nebular spectral models of SN~1998bw. While a similar density structure for SN~2014ad might produce the correct nebular line profiles, any additional steepening of the density at low velocity will increase the mass estimate. Thus, such an explanation cannot reconcile our prediction of unphysically high total mass. \citet{Maeda2006} also found that their nebular spectral models could be improved about equally well by using a two-dimensional model with highly aspherical ejecta. This may be evidence that 2D modeling of SN~2014ad as aspherical is needed to fully reconcile the photospheric and nebular models.

Thus, the predictions from our \tardis\ models can be separately reconciled with either the light-curve analysis or the nebular line fitting, but not both simultaneously. We have assumed that the nebular lines are tracing just the density profile, while emission remains uniform. This assumption may be incorrect, and the peaky nebular features may be caused by a source of extra emission in the core, such as continuing energy from a magnetar. In this case, the agreement in $\alpha$ between our \tardis\ models and the nebular fits is a coincidence, and our \tardis\ models can be reconciled with the light-curve as described above. This seems more likely, but in that case it remains somewhat puzzling that the [O I], [Ca II], and Mg I] have similar line profiles and thus similar nebular fit parameters, as the excitation and emission might be expected to vary differently for these species. In all cases, our analysis suggests that the mass is likely somewhat larger than previously estimated for SN~2014ad.

Degeneracy between the density profile and other \tardis\ parameters, like the abundances, time from explosion, photospheric velocity, and luminosity, could be another possible explanation for the mass discrepancies we find. Though we were unable to find combinations of parameters that gave good spectral fits to both \textit{HST} epochs with lower density, we used relatively simplistic profiles and assumptions. Our assumption of a single power-law density profile works well for times and velocities near the range that our models probe. However, the full density profile likely has a more complicated structure (e.g., perhaps shallower at lower velocities and steeper at higher velocities). Similarly, our assumption of a uniform composition is not realistic, though it does well at epochs (and thus velocities) near the spectra that we model. In attempting to fit later epochs, we find that adjustments to abundance structure at low velocities would be required. Additionally, altering the abundance structure at higher velocities ($>$ 50,000\kms) might improve the \tardis\ model fit in the UV.

Fully exploring more complicated density and abundance profiles for SN~2014ad would likely improve our models; however, these parameters are somewhat degenerate with each other and such exploration by hand may be computationally intractable, given the large parameter space. To address this issue, work is ongoing within the \tardis\ team to study parameter covariance through the use of machine learning and emulators that can vastly speed up model evaluation. For example, a \tardis\ emulator for SN~Ia was developed by \citet{Kerzendorf2021} and \citet{O'Brien2021} with a speedup of several orders of magnitude, and efforts are underway to apply this approach to stripped-envelope SNe (Williamson et al., in preparation). Future work with emulators and posterior distributions may help untangle the degeneracy and provide a path forward. Additional next steps will include detailed modeling of the full UV through IR spectra.

Further limitations of our models arise from the \tardis\ assumption of spherical symmetry. Indeed, the spectropolarimetry from \citet{Stevance2017} suggests that SN~2014ad has large clumps of oxygen and calcium that are asymmetrically distributed in distinct regions from each other, with the calcium layer being deeper in the ejecta. They find evidence of axial symmetry at early epochs and spherical symmetry at late epochs. Jet-like asymmetry in the ejecta of SN Ic-bl can lead to viewing angle effects on the spectral features, as shown by \cite{Barnes_2018}. These spectral differences from polar vs. equatorial viewing angles are relatively minor near maximum-light (where our epochs probe); thus, we would expect that effects from asymmetry might necessitate modification of abundances or velocities somewhat, but would not require drastic changes to the \tardis\ model. At late times and low velocities the ejecta of SN~2014ad is nearly spherical \citep{Stevance2017} and the assumption of spherical symmetry is reasonable. However, asymmetries at early epochs and high velocities might affect the high densities required by our models, and thus impact our estimations of total mass. To account for asymmetric effects, 2-D or 3-D modeling would be required, which is beyond the scope of \tardis\ and this work.

Broadening out to compare SN~2014ad with other relativistic SN Ic-bl without GRBs, we compare our results to the analysis and modeling of SN~2012ap by \citet{Milisavljevic2015}. Our models of SN~2014ad prefer higher photospheric velocities than the \textsc{SYN++} models of SN~2012ap. Our models also imply a higher ejecta mass and kinetic energy than those estimated for SN~2012ap, which are similar in value to the ejecta mass and kinetic energy for SN~2014ad from \citet{Sahu2018} and were calculated in the same way. In agreement with the \textsc{SYN++} modeling of SN~2012ap by \citet{Milisavljevic2015}, in our \tardis\ models we find that O I, Ca II, Fe II, and Si II are crucial contributors to the spectra; however, in contrast, we do not find that He I, C II, or Na I make significant contributions to our \tardis\ models. \citet{Milisavljevic2015} detect high-velocity helium at $+14$d in SN~2012ap, which may agree with the appearance of broad He absorption in the NIR spectra of SN~2014ad from \citet{Shahbandeh_2022} sometime between $+7$d and $+35$d. The nebular [O I] lines in SN~2014ad are similar in shape to SN~2009bb, the first relativistic SN Ibc without a detected GRB \citep{Soderberg2010}. The [O I] profiles are less double-peaked than SN~2012ap, suggesting SN~2014ad may be more spherical at late times than SN~2012ap. The nebular [Ca II] lines in SN~2014ad are symmetric, unlike that of SN~2012ap whose asymmetric [Ca II] was interpreted to be due to opaque interior ejecta \citep{Milisavljevic2015}. Other more recent engine-driven SN Ic-bl without GRBs include iPTF17cw \citep{Corsi2017} and SN 2018bvw (ZTF18aaqjovh) \citep{Ho2020b}, both of which exhibit lower velocities than SN~2014ad. SN~2014ad appears to be another transition object between ``ordinary" SNe Ic-bl and SN-GRBs.

Finally, to pull back to a bigger picture view, in \autoref{fig:sn_uv_diversity} we place the UV spectra of SN~2014ad in context with the UV spectra of other type I supernovae. We include the UV spectra of a SN Ia, SN~2011fe \citep{Mazzali2014}; a type I super-luminous supernova (SLSN I), Gaia\,16apd \citep{Nicholl2017,Yan2018}; a peculiar SN Ib, SN 2006jc \citep{Bufano2009}; the SNe Ic discussed previously in this work, SN~1994I and PTF\,12gzk; and SN Ic-bl, SN~2014ad. Of these comparison objects, the spectra of the SNe Ic are most similar to the SN Ia, in the UV, likely a result of the similar UV line-blanketing from Fe-group elements. Our results show that blue and UV data are powerful diagnostics of stripped envelope supernovae, showing a clear need for continued observations in the future.

\acknowledgments
We thank D.~Sahu, H.~Stevance, S.~Ben-Ami and A.~Gal-Yam for providing access to published data and R.~Fesen for sharing SALT observations. We also thank the anonymous referee whose suggestions significantly improved the paper.

This research has been supported at Rutgers University by NASA/STScI grant HST-GO-13351.001 and NSF award AST-1615455. The SALT data presented here were obtained through Rutgers University programs 2013-2-RU-002 and 2014-1-MLT-001 (PI: Jha) and Dartmouth College program 2013-2-DC\_RSA-001 (PI: Fesen).

M.W. acknowledges support from the NASA Future Investigators in NASA Earth and Space Science and Technology grant (80NSSC21K1849) and support from the Thomas J. Moore Fellowship at New York University.

M.M. was supported in part by the NASA grants 80NSSC21K0240 and 80NSSC22K0486.

K.M. acknowledges support from the JSPS KAKENHI grant (JP18H05223,
JP20H00174, and JP20H04737). 

D. M. acknowledges NSF support from grants PHY-1914448 and AST-2037297.

The UCSC team is supported in part by NSF grants AST-1518052 and AST-1815935, the Gordon \& Betty Moore Foundation, the Heising-Simons Foundation, and from fellowships from the Alfred P.\ Sloan Foundation and the David and Lucile Packard Foundation to R.J.F.

This work made use of \tardis, a community-developed software package for spectral synthesis in supernovae \citep{Kerzendorf2014, kerzendorf_wolfgang_2021_5034859}. The development of \tardis\ received support from GitHub, the Google Summer of Code initiative, and from ESA's Summer of Code in Space program. \tardis\ is a fiscally sponsored project of NumFOCUS. \tardis\ makes extensive use of Astropy and Pyne.

The LBT is an international collaboration among institutions in the United States, Italy and Germany. LBT Corporation partners are: The University of Arizona on behalf of the Arizona Board of Regents; Istituto Nazionale di Astrofisica, Italy; LBT Beteiligungsgesellschaft, Germany, representing the Max-Planck Society, The Leibniz Institute for Astrophysics Potsdam, and Heidelberg University; The Ohio State University, representing OSU, University of Notre Dame, University of Minnesota and University of Virginia.

Some of the data presented in this paper were obtained from the Mikulski Archive for Space Telescopes (MAST) at the Space Telescope Science Institute. The specific observations analyzed can be accessed via \dataset[10.17909/3ajj-4130]{https://doi.org/10.17909/3ajj-4130}.

%\vspace{5mm}
\facilities{HST (STIS), LBT (MODS), Magellan:Baade (IMACS), SALT (RSS)}


\software{\tardis\ \citep{Kerzendorf2014,kerzendorf_wolfgang_2021_5034859}, Astropy \citep{AstropyCollaboration2013, AstropyCollaboration2018}, IRAF \citep{IRAF1, IRAF2}, Matplotlib \citep{Hunter2007}, NumPy \citep{harris2020array}, pandas \citep{McKinney2010}, SciPy \citep{2020SciPy}, Pyraf \citep{Pyraf}, PySALT \citep{PySALT}}




\bibliography{references}{}

\begin{thebibliography}{}
\expandafter\ifx\csname natexlab\endcsname\relax\def\natexlab#1{#1}\fi
\providecommand{\url}[1]{\href{#1}{#1}}
\providecommand{\dodoi}[1]{doi:~\href{http://doi.org/#1}{\nolinkurl{#1}}}
\providecommand{\doeprint}[1]{\href{http://ascl.net/#1}{\nolinkurl{http://ascl.net/#1}}}
\providecommand{\doarXiv}[1]{\href{https://arxiv.org/abs/#1}{\nolinkurl{https://arxiv.org/abs/#1}}}

\bibitem[{{Arnett}(1982)}]{Arnett1982}
{Arnett}, W.~D. 1982, \apj, 253, 785, \dodoi{10.1086/159681}

\bibitem[{{Astropy Collaboration} {et~al.}(2013){Astropy Collaboration},
  {Robitaille}, {Tollerud}, {Greenfield}, {Droettboom}, {Bray}, {Aldcroft},
  {Davis}, {Ginsburg}, {Price-Whelan}, {Kerzendorf}, {Conley}, {Crighton},
  {Barbary}, {Muna}, {Ferguson}, {Grollier}, {Parikh}, {Nair}, {Unther},
  {Deil}, {Woillez}, {Conseil}, {Kramer}, {Turner}, {Singer}, {Fox}, {Weaver},
  {Zabalza}, {Edwards}, {Azalee Bostroem}, {Burke}, {Casey}, {Crawford},
  {Dencheva}, {Ely}, {Jenness}, {Labrie}, {Lim}, {Pierfederici}, {Pontzen},
  {Ptak}, {Refsdal}, {Servillat}, \& {Streicher}}]{AstropyCollaboration2013}
{Astropy Collaboration}, {Robitaille}, T.~P., {Tollerud}, E.~J., {et~al.} 2013,
  \aap, 558, A33, \dodoi{10.1051/0004-6361/201322068}

\bibitem[{{Astropy Collaboration} {et~al.}(2018){Astropy Collaboration},
  {Price-Whelan}, {Sip{\H{o}}cz}, {G{\"u}nther}, {Lim}, {Crawford}, {Conseil},
  {Shupe}, {Craig}, {Dencheva}, {Ginsburg}, {VanderPlas}, {Bradley},
  {P{\'e}rez-Su{\'a}rez}, {de Val-Borro}, {Aldcroft}, {Cruz}, {Robitaille},
  {Tollerud}, {Ardelean}, {Babej}, {Bach}, {Bachetti}, {Bakanov}, {Bamford},
  {Barentsen}, {Barmby}, {Baumbach}, {Berry}, {Biscani}, {Boquien}, {Bostroem},
  {Bouma}, {Brammer}, {Bray}, {Breytenbach}, {Buddelmeijer}, {Burke},
  {Calderone}, {Cano Rodr{\'\i}guez}, {Cara}, {Cardoso}, {Cheedella}, {Copin},
  {Corrales}, {Crichton}, {D'Avella}, {Deil}, {Depagne}, {Dietrich}, {Donath},
  {Droettboom}, {Earl}, {Erben}, {Fabbro}, {Ferreira}, {Finethy}, {Fox},
  {Garrison}, {Gibbons}, {Goldstein}, {Gommers}, {Greco}, {Greenfield},
  {Groener}, {Grollier}, {Hagen}, {Hirst}, {Homeier}, {Horton}, {Hosseinzadeh},
  {Hu}, {Hunkeler}, {Ivezi{\'c}}, {Jain}, {Jenness}, {Kanarek}, {Kendrew},
  {Kern}, {Kerzendorf}, {Khvalko}, {King}, {Kirkby}, {Kulkarni}, {Kumar},
  {Lee}, {Lenz}, {Littlefair}, {Ma}, {Macleod}, {Mastropietro}, {McCully},
  {Montagnac}, {Morris}, {Mueller}, {Mumford}, {Muna}, {Murphy}, {Nelson},
  {Nguyen}, {Ninan}, {N{\"o}the}, {Ogaz}, {Oh}, {Parejko}, {Parley}, {Pascual},
  {Patil}, {Patil}, {Plunkett}, {Prochaska}, {Rastogi}, {Reddy Janga},
  {Sabater}, {Sakurikar}, {Seifert}, {Sherbert}, {Sherwood-Taylor}, {Shih},
  {Sick}, {Silbiger}, {Singanamalla}, {Singer}, {Sladen}, {Sooley},
  {Sornarajah}, {Streicher}, {Teuben}, {Thomas}, {Tremblay}, {Turner},
  {Terr{\'o}n}, {van Kerkwijk}, {de la Vega}, {Watkins}, {Weaver}, {Whitmore},
  {Woillez}, {Zabalza}, \& {Astropy Contributors}}]{AstropyCollaboration2018}
{Astropy Collaboration}, {Price-Whelan}, A.~M., {Sip{\H{o}}cz}, B.~M., {et~al.}
  2018, \aj, 156, 123, \dodoi{10.3847/1538-3881/aabc4f}

\bibitem[{{Barna} {et~al.}(2021){Barna}, {Pereira}, {Taubenberger}, {Magee},
  {Kromer}, {Kerzendorf}, {Vogl}, {Williamson}, {Fl{\"o}rs}, {Noebauer},
  {Foley}, {Sasdelli}, \& {Hillebrandt}}]{Barna2021}
{Barna}, B., {Pereira}, T., {Taubenberger}, S., {et~al.} 2021, \mnras,
  \dodoi{10.1093/mnras/stab1736}

\bibitem[{Barnes {et~al.}(2018)Barnes, Duffell, Liu, Modjaz, Bianco, Kasen, \&
  MacFadyen}]{Barnes_2018}
Barnes, J., Duffell, P.~C., Liu, Y., {et~al.} 2018, The Astrophysical Journal,
  860, 38, \dodoi{10.3847/1538-4357/aabf84}

\bibitem[{{Ben-Ami} {et~al.}(2012){Ben-Ami}, {Gal-Yam}, {Filippenko},
  {Mazzali}, {Modjaz}, {Yaron}, {Arcavi}, {Cenko}, {Horesh}, {Howell},
  {Graham}, {Horst}, {Im}, {Jeon}, {Kulkarni}, {Leonard}, {Perley}, {Pian},
  {Sand}, {Sullivan}, {Becker}, {Bersier}, {Bloom}, {Bottom}, {Brown}, {Clubb},
  {Dilday}, {Dixon}, {Fortinsky}, {Fox}, {Gonzalez}, {Harutyunyan}, {Kasliwal},
  {Li}, {Malkan}, {Manulis}, {Matheson}, {Moskovitz}, {Muirhead}, {Nugent},
  {Ofek}, {Quimby}, {Richards}, {Ross}, {Searcy}, {Silverman}, {Smith},
  {Vanderburg}, \& {Walker}}]{Ben-Ami2012}
{Ben-Ami}, S., {Gal-Yam}, A., {Filippenko}, A.~V., {et~al.} 2012, \apjl, 760,
  L33, \dodoi{10.1088/2041-8205/760/2/L33}

\bibitem[{{Ben-Ami} {et~al.}(2015){Ben-Ami}, {Hachinger}, {Gal-Yam}, {Mazzali},
  {Filippenko}, {Horesh}, {Matheson}, {Modjaz}, {Sauer}, {Silverman}, {Smith},
  \& {Yaron}}]{Ben-Ami2015}
{Ben-Ami}, S., {Hachinger}, S., {Gal-Yam}, A., {et~al.} 2015, \apj, 803, 40,
  \dodoi{10.1088/0004-637X/803/1/40}

\bibitem[{{Bufano} {et~al.}(2009){Bufano}, {Immler}, {Turatto}, {Landsman},
  {Brown}, {Benetti}, {Cappellaro}, {Holland}, {Mazzali}, {Milne}, {Panagia},
  {Pian}, {Roming}, {Zampieri}, {Breeveld}, \& {Gehrels}}]{Bufano2009}
{Bufano}, F., {Immler}, S., {Turatto}, M., {et~al.} 2009, \apj, 700, 1456,
  \dodoi{10.1088/0004-637X/700/2/1456}

\bibitem[{Cano {et~al.}(2017)Cano, Wang, Dai, \& Wu}]{Cano_2017}
Cano, Z., Wang, S.-Q., Dai, Z.-G., \& Wu, X.-F. 2017, Advances in Astronomy,
  2017, 1–41, \dodoi{10.1155/2017/8929054}

\bibitem[{{Chornock} {et~al.}(2013){Chornock}, {Berger}, {Rest},
  {Milisavljevic}, {Lunnan}, {Foley}, {Soderberg}, {Smartt}, {Burgasser},
  {Challis}, {Chomiuk}, {Czekala}, {Drout}, {Fong}, {Huber}, {Kirshner},
  {Leibler}, {McLeod}, {Marion}, {Narayan}, {Riess}, {Roth}, {Sanders},
  {Scolnic}, {Smith}, {Stubbs}, {Tonry}, {Valenti}, {Burgett}, {Chambers},
  {Hodapp}, {Kaiser}, {Kudritzki}, {Magnier}, \& {Price}}]{Chornock:2013}
{Chornock}, R., {Berger}, E., {Rest}, A., {et~al.} 2013, \apj, 767, 162,
  \dodoi{10.1088/0004-637X/767/2/162}

\bibitem[{{Clocchiatti} \& {Wheeler}(1997)}]{Clocchiatti1997}
{Clocchiatti}, A., \& {Wheeler}, J.~C. 1997, \apj, 491, 375,
  \dodoi{10.1086/304961}

\bibitem[{Corsi {et~al.}(2016)Corsi, Gal-Yam, Kulkarni, Frail, Mazzali, Cenko,
  Kasliwal, Cao, Horesh, Palliyaguru, Perley, Laher, Taddia, Leloudas, Maguire,
  Nugent, Sollerman, \& Sullivan}]{Corsi_2016}
Corsi, A., Gal-Yam, A., Kulkarni, S.~R., {et~al.} 2016, The Astrophysical
  Journal, 830, 42, \dodoi{10.3847/0004-637x/830/1/42}

\bibitem[{{Corsi} {et~al.}(2017){Corsi}, {Cenko}, {Kasliwal}, {Quimby},
  {Kulkarni}, {Frail}, {Goldstein}, {Blagorodnova}, {Connaughton}, {Perley},
  {Singer}, {Copperwheat}, {Fremling}, {Kupfer}, {Piascik}, {Steele}, {Taddia},
  {Vedantham}, {Kutyrev}, {Palliyaguru}, {Roberts}, {Sollerman}, {Troja}, \&
  {Veilleux}}]{Corsi2017}
{Corsi}, A., {Cenko}, S.~B., {Kasliwal}, M.~M., {et~al.} 2017, \apj, 847, 54,
  \dodoi{10.3847/1538-4357/aa85e5}

\bibitem[{{Crawford} {et~al.}(2010){Crawford}, {Still}, {Schellart}, {Balona},
  {Buckley}, {Dugmore}, {Gulbis}, {Kniazev}, {Kotze}, {Loaring}, {Nordsieck},
  {Pickering}, {Potter}, {Romero Colmenero}, {Vaisanen}, {Williams}, \&
  {Zietsman}}]{PySALT}
{Crawford}, S.~M., {Still}, M., {Schellart}, P., {et~al.} 2010, in Society of
  Photo-Optical Instrumentation Engineers (SPIE) Conference Series, Vol. 7737,
  Society of Photo-Optical Instrumentation Engineers (SPIE) Conference Series,
  25, \dodoi{10.1117/12.857000}

\bibitem[{{Crowther}(2007)}]{Crowther2007}
{Crowther}, P.~A. 2007, \araa, 45, 177,
  \dodoi{10.1146/annurev.astro.45.051806.110615}

\bibitem[{{Dessart} {et~al.}(2012){Dessart}, {Hillier}, {Li}, \&
  {Woosley}}]{Dessart2012}
{Dessart}, L., {Hillier}, D.~J., {Li}, C., \& {Woosley}, S. 2012, \mnras, 424,
  2139, \dodoi{10.1111/j.1365-2966.2012.21374.x}

\bibitem[{{Dessart} {et~al.}(2015){Dessart}, {Hillier}, {Woosley}, {Livne},
  {Waldman}, {Yoon}, \& {Langer}}]{Dessart2015}
{Dessart}, L., {Hillier}, D.~J., {Woosley}, S., {et~al.} 2015, \mnras, 453,
  2189, \dodoi{10.1093/mnras/stv1747}

\bibitem[{{Djorgovski} {et~al.}(2011){Djorgovski}, {Donalek}, {Mahabal},
  {Moghaddam}, {Turmon}, {Graham}, {Drake}, {Sharma}, \&
  {Chen}}]{Djorgovski2011}
{Djorgovski}, S.~G., {Donalek}, C., {Mahabal}, A., {et~al.} 2011, arXiv
  e-prints, arXiv:1110.4655.
\newblock \doarXiv{1110.4655}

\bibitem[{{Dressler} {et~al.}(2011){Dressler}, {Bigelow}, {Hare}, {Sutin},
  {Thompson}, {Burley}, {Epps}, {Oemler}, {Bagish}, {Birk}, {Clardy},
  {Gunnels}, {Kelson}, {Shectman}, \& {Osip}}]{Dressler11}
{Dressler}, A., {Bigelow}, B., {Hare}, T., {et~al.} 2011, \pasp, 123, 288,
  \dodoi{10.1086/658908}

\bibitem[{{Filippenko} {et~al.}(1993){Filippenko}, {Matheson}, \&
  {Ho}}]{Filippenko1993}
{Filippenko}, A.~V., {Matheson}, T., \& {Ho}, L.~C. 1993, \apjl, 415, L103,
  \dodoi{10.1086/187043}

\bibitem[{{Filippenko} {et~al.}(1995){Filippenko}, {Barth}, {Matheson},
  {Armus}, {Brown}, {Espey}, {Fan}, {Goodrich}, {Ho}, {Junkkarinen}, {Koo},
  {Lehnert}, {Martel}, {Mazzarella}, {Miller}, {Smith}, {Tytler}, \&
  {Wirth}}]{Filippenko1995}
{Filippenko}, A.~V., {Barth}, A.~J., {Matheson}, T., {et~al.} 1995, \apjl, 450,
  L11, \dodoi{10.1086/309659}

\bibitem[{{Fisher}(2000)}]{Fisher2000}
{Fisher}, A.~K. 2000, PhD thesis, University of Oklahoma, United States

\bibitem[{{Fitzpatrick}(1999)}]{Fitzpatrick1999}
{Fitzpatrick}, E.~L. 1999, \pasp, 111, 63, \dodoi{10.1086/316293}

\bibitem[{{Foley} {et~al.}(2020){Foley}, {Dimitriadis}, {Filippenko}, {Fox},
  {Jones}, {Kirshner}, {Pan}, {Roepke}, {Rojas-Bravo}, {Siebert}, {Sim},
  {Taubenberger}, \& {Wheeler}}]{Foley2020}
{Foley}, R., {Dimitriadis}, G., {Filippenko}, A.~V., {et~al.} 2020, {Measuring
  the Effect of Progenitor Metallicity on Type Ia Supernova Distance
  Estimates}, HST Proposal. Cycle 28, ID. \#16238

\bibitem[{{Foley} {et~al.}(2012){Foley}, {Kromer}, {Howie Marion}, {Pignata},
  {Stritzinger}, {Taubenberger}, {Challis}, {Filippenko}, {Folatelli},
  {Hillebrandt}, {Hsiao}, {Kirshner}, {Li}, {Morrell}, {R{\"o}pke},
  {Ciaraldi-Schoolmann}, {Seitenzahl}, {Silverman}, {Simcoe}, {Berta},
  {Ivarsen}, {Newton}, {Nysewander}, \& {Reichart}}]{Foley2012}
{Foley}, R.~J., {Kromer}, M., {Howie Marion}, G., {et~al.} 2012, \apjl, 753,
  L5, \dodoi{10.1088/2041-8205/753/1/L5}

\bibitem[{{Foley} {et~al.}(2014){Foley}, {Fox}, {McCully}, {Phillips}, {Sand},
  {Zheng}, {Challis}, {Filippenko}, {Folatelli}, {Hillebrandt}, {Hsiao}, {Jha},
  {Kirshner}, {Kromer}, {Marion}, {Nelson}, {Pakmor}, {Pignata}, {R{\"o}pke},
  {Seitenzahl}, {Silverman}, {Skrutskie}, \& {Stritzinger}}]{Foley2014}
{Foley}, R.~J., {Fox}, O.~D., {McCully}, C., {et~al.} 2014, \mnras, 443, 2887,
  \dodoi{10.1093/mnras/stu1378}

\bibitem[{{Foley} {et~al.}(2016){Foley}, {Pan}, {Brown}, {Filippenko}, {Fox},
  {Hillebrandt}, {Kirshner}, {Marion}, {Milne}, {Parrent}, {Pignata}, \&
  {Stritzinger}}]{Foley2016}
{Foley}, R.~J., {Pan}, Y.-C., {Brown}, P., {et~al.} 2016, \mnras, 461, 1308,
  \dodoi{10.1093/mnras/stw1440}

\bibitem[{{Foreman-Mackey} {et~al.}(2013){Foreman-Mackey}, {Hogg}, {Lang}, \&
  {Goodman}}]{emcee}
{Foreman-Mackey}, D., {Hogg}, D.~W., {Lang}, D., \& {Goodman}, J. 2013, \pasp,
  125, 306, \dodoi{10.1086/670067}

\bibitem[{Gal-Yam(2017)}]{Gal-Yam2017}
Gal-Yam, A. 2017, Observational and Physical Classification of Supernovae, ed.
  A.~W. Alsabti \& P.~Murdin (Cham: Springer International Publishing),
  195--237, \dodoi{10.1007/978-3-319-21846-5_35}

\bibitem[{Harris {et~al.}(2020)Harris, Millman, van~der Walt, Gommers,
  Virtanen, Cournapeau, Wieser, Taylor, Berg, Smith, Kern, Picus, Hoyer, van
  Kerkwijk, Brett, Haldane, del R{\'{i}}o, Wiebe, Peterson,
  G{\'{e}}rard-Marchant, Sheppard, Reddy, Weckesser, Abbasi, Gohlke, \&
  Oliphant}]{harris2020array}
Harris, C.~R., Millman, K.~J., van~der Walt, S.~J., {et~al.} 2020, Nature, 585,
  357, \dodoi{10.1038/s41586-020-2649-2}

\bibitem[{{Heger} {et~al.}(2003){Heger}, {Fryer}, {Woosley}, {Langer}, \&
  {Hartmann}}]{Heger2003}
{Heger}, A., {Fryer}, C.~L., {Woosley}, S.~E., {Langer}, N., \& {Hartmann},
  D.~H. 2003, \apj, 591, 288, \dodoi{10.1086/375341}

\bibitem[{{Ho} {et~al.}(2020{\natexlab{a}}){Ho}, {Kulkarni}, {Perley}, {Cenko},
  {Corsi}, {Schulze}, {Lunnan}, {Sollerman}, {Gal-Yam}, {Anand}, {Barbarino},
  {Bellm}, {Bruch}, {Burns}, {De}, {Dekany}, {Delacroix}, {Duev}, {Frederiks},
  {Fremling}, {Goldstein}, {Golkhou}, {Graham}, {Hale}, {Kasliwal}, {Kupfer},
  {Laher}, {Martikainen}, {Masci}, {Neill}, {Ridnaia}, {Rusholme}, {Savchenko},
  {Shupe}, {Soumagnac}, {Strotjohann}, {Svinkin}, {Taggart}, {Tartaglia},
  {Yan}, \& {Zolkower}}]{Ho2020a}
{Ho}, A. Y.~Q., {Kulkarni}, S.~R., {Perley}, D.~A., {et~al.}
  2020{\natexlab{a}}, \apj, 902, 86, \dodoi{10.3847/1538-4357/aba630}

\bibitem[{{Ho} {et~al.}(2020{\natexlab{b}}){Ho}, {Corsi}, {Cenko}, {Taddia},
  {Kulkarni}, {Adams}, {De}, {Dekany}, {Frederiks}, {Fremling}, {Golkhou},
  {Graham}, {Hung}, {Kupfer}, {Laher}, {Mahabal}, {Masci}, {Miller}, {Neill},
  {Reiley}, {Riddle}, {Ridnaia}, {Rusholme}, {Sharma}, {Sollerman},
  {Soumagnac}, {Svinkin}, \& {Shupe}}]{Ho2020b}
{Ho}, A. Y.~Q., {Corsi}, A., {Cenko}, S.~B., {et~al.} 2020{\natexlab{b}}, \apj,
  893, 132, \dodoi{10.3847/1538-4357/ab7f3b}

\bibitem[{{Howerton} {et~al.}(2014){Howerton}, {Drake}, {Djorgovski},
  {Mahabal}, {Graham}, {Williams}, {Prieto}, {Catelan}, {Christensen},
  {Larson}, {Masi}, {Nocentini}, {Schmeer}, {Luppi}, {Buzzi}, {Foglia},
  {Concari}, {Galli}, {Tombelli}, {Jha}, {Pandya}, {McCully}, {Foley},
  {Crause}, \& {Garnavich}}]{Howerton2014}
{Howerton}, S., {Drake}, A.~J., {Djorgovski}, S.~G., {et~al.} 2014, Central
  Bureau Electronic Telegrams, 3831, 1

\bibitem[{{Hunter}(2007)}]{Hunter2007}
{Hunter}, J.~D. 2007, Computing in Science and Engineering, 9, 90,
  \dodoi{10.1109/MCSE.2007.55}

\bibitem[{{Iwamoto} {et~al.}(1998){Iwamoto}, {Mazzali}, {Nomoto}, {Umeda},
  {Nakamura}, {Patat}, {Danziger}, {Young}, {Suzuki}, {Shigeyama},
  {Augusteijn}, {Doublier}, {Gonzalez}, {Boehnhardt}, {Brewer}, {Hainaut},
  {Lidman}, {Leibundgut}, {Cappellaro}, {Turatto}, {Galama}, {Vreeswijk},
  {Kouveliotou}, {van Paradijs}, {Pian}, {Palazzi}, \&
  {Frontera}}]{Iwamoto1998}
{Iwamoto}, K., {Mazzali}, P.~A., {Nomoto}, K., {et~al.} 1998, \nat, 395, 672,
  \dodoi{10.1038/27155}

\bibitem[{{Izzo} {et~al.}(2020){Izzo}, {Auchettl}, {Hjorth}, {De Colle},
  {Gall}, {Angus}, {Raimundo}, \& {Ramirez-Ruiz}}]{Izzo2020}
{Izzo}, L., {Auchettl}, K., {Hjorth}, J., {et~al.} 2020, \aap, 639, L11,
  \dodoi{10.1051/0004-6361/202038152}

\bibitem[{{Izzo} {et~al.}(2019){Izzo}, {de Ugarte Postigo}, {Maeda},
  {Th{\"o}ne}, {Kann}, {Della Valle}, {Sagues Carracedo}, {Micha{\l}owski},
  {Schady}, {Schmidl}, {Selsing}, {Starling}, {Suzuki}, {Bensch}, {Bolmer},
  {Campana}, {Cano}, {Covino}, {Fynbo}, {Hartmann}, {Heintz}, {Hjorth},
  {Japelj}, {Kami{\'n}ski}, {Kaper}, {Kouveliotou}, {Kru{\.Z}y{\'n}ski},
  {Kwiatkowski}, {Leloudas}, {Levan}, {Malesani}, {Micha{\l}owski},
  {Piranomonte}, {Pugliese}, {Rossi}, {S{\'a}nchez-Ram{\'\i}rez}, {Schulze},
  {Steeghs}, {Tanvir}, {Ulaczyk}, {Vergani}, \& {Wiersema}}]{Izzo2019}
{Izzo}, L., {de Ugarte Postigo}, A., {Maeda}, K., {et~al.} 2019, \nat, 565,
  324, \dodoi{10.1038/s41586-018-0826-3}

\bibitem[{{Jerkstrand}(2017)}]{Jerkstrand2017}
{Jerkstrand}, A. 2017, in Handbook of Supernovae, ed. A.~W. {Alsabti} \&
  P.~{Murdin}, 795, \dodoi{10.1007/978-3-319-21846-5\_29}

\bibitem[{{Katsuda} {et~al.}(2018){Katsuda}, {Takiwaki}, {Tominaga}, {Moriya},
  \& {Nakamura}}]{Katsuda2018}
{Katsuda}, S., {Takiwaki}, T., {Tominaga}, N., {Moriya}, T.~J., \& {Nakamura},
  K. 2018, \apj, 863, 127, \dodoi{10.3847/1538-4357/aad2d8}

\bibitem[{Kerzendorf {et~al.}(2021)Kerzendorf, Sim, Vogl, Williamson, Pássaro,
  Flörs, Camacho, Jančauskas, Harpole, Nöbauer, Lietzau, Mishin, Tsamis,
  Boyle, Shingles, Gupta, Desai, Klauser, Beaujean, Suban-Loewen, Heringer,
  Barna, Nayak~U, Gautam, Kumar, Kharkar, Yap, Dasgupta, Volodin, Sandler, k,
  Patra, Singh~Rathore, Patel, Sharma, Gupta, Wahi, Aggarwal, Sarafina,
  Kowalski, Sofiatti, Selsing, Talegaonkar, Yu, Brar, Singh, Jain, Floers,
  Reichenbach, Mishra, Rajagopalan, Magee, Livneh, Arya, Alam, Sondhi, Eguren,
  Bentil, Savel, Bylund, Eweis, Reinecke, O'Brien, Gillanders, Varanasi, Patel,
  Barbosa, Smith, Singhal, Cawley, \&
  Fullard}]{kerzendorf_wolfgang_2021_5034859}
Kerzendorf, W., Sim, S., Vogl, C., {et~al.} 2021, tardis-sn/tardis: TARDIS
  v3.0.dev4019, v3.0.dev4019,  Zenodo, \dodoi{10.5281/zenodo.5034859}

\bibitem[{{Kerzendorf} \& {Sim}(2014)}]{Kerzendorf2014}
{Kerzendorf}, W.~E., \& {Sim}, S.~A. 2014, \mnras, 440, 387,
  \dodoi{10.1093/mnras/stu055}

\bibitem[{{Kerzendorf} {et~al.}(2021){Kerzendorf}, {Vogl}, {Buchner},
  {Contardo}, {Williamson}, \& {van der Smagt}}]{Kerzendorf2021}
{Kerzendorf}, W.~E., {Vogl}, C., {Buchner}, J., {et~al.} 2021, \apjl, 910, L23,
  \dodoi{10.3847/2041-8213/abeb1b}

\bibitem[{{Langer}(2012)}]{Langer2012}
{Langer}, N. 2012, \araa, 50, 107, \dodoi{10.1146/annurev-astro-081811-125534}

\bibitem[{Lazzati {et~al.}(2012)Lazzati, Morsony, Blackwell, \&
  Begelman}]{lazzati2012unifying}
Lazzati, D., Morsony, B.~J., Blackwell, C.~H., \& Begelman, M.~C. 2012, The
  Astrophysical Journal, 750, 68

\bibitem[{{Liu} {et~al.}(2016){Liu}, {Modjaz}, {Bianco}, \& {Graur}}]{Liu2016}
{Liu}, Y.-Q., {Modjaz}, M., {Bianco}, F.~B., \& {Graur}, O. 2016, \apj, 827,
  90, \dodoi{10.3847/0004-637X/827/2/90}

\bibitem[{{Maeda} {et~al.}(2006){Maeda}, {Nomoto}, {Mazzali}, \&
  {Deng}}]{Maeda2006}
{Maeda}, K., {Nomoto}, K., {Mazzali}, P.~A., \& {Deng}, J. 2006, \apj, 640,
  854, \dodoi{10.1086/500187}

\bibitem[{{Maeda} {et~al.}(2007){Maeda}, {Kawabata}, {Tanaka}, {Nomoto},
  {Tominaga}, {Hattori}, {Minezaki}, {Kuroda}, {Suzuki}, {Deng}, {Mazzali}, \&
  {Pian}}]{Maeda2007}
{Maeda}, K., {Kawabata}, K., {Tanaka}, M., {et~al.} 2007, \apjl, 658, L5,
  \dodoi{10.1086/513564}

\bibitem[{{Maeder}(1987)}]{Maeder1987}
{Maeder}, A. 1987, \aap, 178, 159

\bibitem[{{Magee} {et~al.}(2016){Magee}, {Kotak}, {Sim}, {Kromer},
  {Rabinowitz}, {Smartt}, {Baltay}, {Campbell}, {Chen}, {Fink}, {Gal-Yam},
  {Galbany}, {Hillebrandt}, {Inserra}, {Kankare}, {Le Guillou}, {Lyman},
  {Maguire}, {Pakmor}, {R{\"o}pke}, {Ruiter}, {Seitenzahl}, {Sullivan},
  {Valenti}, \& {Young}}]{Magee2016}
{Magee}, M.~R., {Kotak}, R., {Sim}, S.~A., {et~al.} 2016, \aap, 589, A89,
  \dodoi{10.1051/0004-6361/201528036}

\bibitem[{{Margutti} {et~al.}(2014){Margutti}, {Milisavljevic}, {Soderberg},
  {Guidorzi}, {Morsony}, {Sanders}, {Chakraborti}, {Ray}, {Kamble}, {Drout},
  {Parrent}, {Zauderer}, \& {Chomiuk}}]{Margutti2014}
{Margutti}, R., {Milisavljevic}, D., {Soderberg}, A.~M., {et~al.} 2014, \apj,
  797, 107, \dodoi{10.1088/0004-637X/797/2/107}

\bibitem[{{Marongiu} {et~al.}(2019){Marongiu}, {Guidorzi}, {Margutti},
  {Coppejans}, {Martone}, \& {Kamble}}]{Marongiu2019}
{Marongiu}, M., {Guidorzi}, C., {Margutti}, R., {et~al.} 2019, \apj, 879, 89,
  \dodoi{10.3847/1538-4357/ab25ef}

\bibitem[{{Mazzali} {et~al.}(2000){Mazzali}, {Iwamoto}, \&
  {Nomoto}}]{Mazzali2000}
{Mazzali}, P.~A., {Iwamoto}, K., \& {Nomoto}, K. 2000, \apj, 545, 407,
  \dodoi{10.1086/317808}

\bibitem[{{Mazzali} {et~al.}(2021){Mazzali}, {Pian}, {Bufano}, \&
  {Ashall}}]{Mazzali21}
{Mazzali}, P.~A., {Pian}, E., {Bufano}, F., \& {Ashall}, C. 2021, \mnras, 505,
  4106, \dodoi{10.1093/mnras/stab1594}

\bibitem[{{Mazzali} {et~al.}(2002){Mazzali}, {Deng}, {Maeda}, {Nomoto},
  {Umeda}, {Hatano}, {Iwamoto}, {Yoshii}, {Kobayashi}, {Minezaki}, {Doi},
  {Enya}, {Tomita}, {Smartt}, {Kinugasa}, {Kawakita}, {Ayani}, {Kawabata},
  {Yamaoka}, {Qiu}, {Motohara}, {Gerardy}, {Fesen}, {Kawabata}, {Iye},
  {Kashikawa}, {Kosugi}, {Ohyama}, {Takada-Hidai}, {Zhao}, {Chornock},
  {Filippenko}, {Benetti}, \& {Turatto}}]{Mazzali2002}
{Mazzali}, P.~A., {Deng}, J., {Maeda}, K., {et~al.} 2002, \apjl, 572, L61,
  \dodoi{10.1086/341504}

\bibitem[{{Mazzali} {et~al.}(2014){Mazzali}, {Sullivan}, {Hachinger}, {Ellis},
  {Nugent}, {Howell}, {Gal-Yam}, {Maguire}, {Cooke}, {Thomas}, {Nomoto}, \&
  {Walker}}]{Mazzali2014}
{Mazzali}, P.~A., {Sullivan}, M., {Hachinger}, S., {et~al.} 2014, \mnras, 439,
  1959, \dodoi{10.1093/mnras/stu077}

\bibitem[{McKinney(2010)}]{McKinney2010}
McKinney, W. 2010, in Proceedings of the 9th Python in Science Conference, ed.
  S.~van~der Walt \& J.~Millman, 51 -- 56

\bibitem[{{Milisavljevic} {et~al.}(2010){Milisavljevic}, {Fesen}, {Gerardy},
  {Kirshner}, \& {Challis}}]{Milisavljevic2010}
{Milisavljevic}, D., {Fesen}, R.~A., {Gerardy}, C.~L., {Kirshner}, R.~P., \&
  {Challis}, P. 2010, \apj, 709, 1343, \dodoi{10.1088/0004-637X/709/2/1343}

\bibitem[{{Milisavljevic} {et~al.}(2015){Milisavljevic}, {Margutti}, {Parrent},
  {Soderberg}, {Fesen}, {Mazzali}, {Maeda}, {Sanders}, {Cenko}, {Silverman},
  {Filippenko}, {Kamble}, {Chakraborti}, {Drout}, {Kirshner}, {Pickering},
  {Kawabata}, {Hattori}, {Hsiao}, {Stritzinger}, {Marion}, {Vinko}, \&
  {Wheeler}}]{Milisavljevic2015}
{Milisavljevic}, D., {Margutti}, R., {Parrent}, J.~T., {et~al.} 2015, \apj,
  799, 51, \dodoi{10.1088/0004-637X/799/1/51}

\bibitem[{{Miller} {et~al.}(2021){Miller}, {Cenko}, {Gal-Yam}, {Goobar},
  {Graham}, {Kasliwal}, {Kulkarni}, {Maguire}, {Nugent}, {Ofek}, \&
  {Sollerman}}]{Miller2021}
{Miller}, A., {Cenko}, S.~B., {Gal-Yam}, A., {et~al.} 2021, {Caught in the Act:
  UV spectroscopy of the ejecta-companion collision from a type Ia supernova},
  HST Proposal. Cycle 29, ID. \#16732

\bibitem[{{Modjaz}(2011)}]{Modjaz11}
{Modjaz}, M. 2011, Astronomische Nachrichten, 332, 434,
  \dodoi{10.1002/asna.201111562}

\bibitem[{{Modjaz} {et~al.}(2019){Modjaz}, {Guti{\'e}rrez}, \&
  {Arcavi}}]{Modjaz19}
{Modjaz}, M., {Guti{\'e}rrez}, C.~P., \& {Arcavi}, I. 2019, Nature Astronomy,
  3, 717, \dodoi{10.1038/s41550-019-0856-2}

\bibitem[{{Modjaz} {et~al.}(2008){Modjaz}, {Kirshner}, {Blondin}, {Challis}, \&
  {Matheson}}]{Modjaz08}
{Modjaz}, M., {Kirshner}, R.~P., {Blondin}, S., {Challis}, P., \& {Matheson},
  T. 2008, \apjl, 687, L9, \dodoi{10.1086/593135}

\bibitem[{Modjaz {et~al.}(2016)Modjaz, Liu, Bianco, \& Graur}]{mlb16}
Modjaz, M., Liu, Y.~Q., Bianco, F.~B., \& Graur, O. 2016, The Astrophysical
  Journal, 832, 108, \dodoi{10.3847/0004-637x/832/2/108}

\bibitem[{{Modjaz} {et~al.}(2016){Modjaz}, {Liu}, {Bianco}, \&
  {Graur}}]{Modjaz2016}
{Modjaz}, M., {Liu}, Y.~Q., {Bianco}, F.~B., \& {Graur}, O. 2016, \apj, 832,
  108, \dodoi{10.3847/0004-637X/832/2/108}

\bibitem[{{Modjaz} {et~al.}(2020){Modjaz}, {Bianco}, {Siwek}, {Huang},
  {Perley}, {Fierroz}, {Liu}, {Arcavi}, {Gal-Yam}, {Filippenko},
  {Blagorodnova}, {Cenko}, {Kasliwal}, {Kulkarni}, {Schulze}, {Taggart}, \&
  {Zheng}}]{Modjaz20}
{Modjaz}, M., {Bianco}, F.~B., {Siwek}, M., {et~al.} 2020, \apj, 892, 153,
  \dodoi{10.3847/1538-4357/ab4185}

\bibitem[{{Nicholl} {et~al.}(2017){Nicholl}, {Berger}, {Margutti}, {Blanchard},
  {Milisavljevic}, {Challis}, {Metzger}, \& {Chornock}}]{Nicholl2017}
{Nicholl}, M., {Berger}, E., {Margutti}, R., {et~al.} 2017, \apjl, 835, L8,
  \dodoi{10.3847/2041-8213/aa56c5}

\bibitem[{{Nomoto} {et~al.}(2006){Nomoto}, {Tominaga}, {Umeda}, {Kobayashi}, \&
  {Maeda}}]{Nomoto2006}
{Nomoto}, K., {Tominaga}, N., {Umeda}, H., {Kobayashi}, C., \& {Maeda}, K.
  2006, \nphysa, 777, 424, \dodoi{10.1016/j.nuclphysa.2006.05.008}

\bibitem[{{O'Brien} {et~al.}(2021){O'Brien}, {Kerzendorf}, {Fullard},
  {Williamson}, {Pakmor}, {Buchner}, {Hachinger}, {Vogl}, {Gillanders},
  {Fl{\"o}rs}, \& {van der Smagt}}]{O'Brien2021}
{O'Brien}, J.~T., {Kerzendorf}, W.~E., {Fullard}, A., {et~al.} 2021, \apjl,
  916, L14, \dodoi{10.3847/2041-8213/ac1173}

\bibitem[{{Pettini} \& {Pagel}(2004)}]{Pettini04}
{Pettini}, M., \& {Pagel}, B. E.~J. 2004, \mnras, 348, L59,
  \dodoi{10.1111/j.1365-2966.2004.07591.x}

\bibitem[{{Podsiadlowski} {et~al.}(1992){Podsiadlowski}, {Joss}, \&
  {Hsu}}]{Podsiadlowski1992}
{Podsiadlowski}, P., {Joss}, P.~C., \& {Hsu}, J.~J.~L. 1992, \apj, 391, 246,
  \dodoi{10.1086/171341}

\bibitem[{{Pogge} {et~al.}(2012){Pogge}, {Atwood}, {O'Brien}, {Byard},
  {Derwent}, {Gonzalez}, {Martini}, {Mason}, {Osmer}, {Pappalardo}, {Zhelem},
  {Stoll}, {Steinbrecher}, {Brewer}, {Colarosa}, \& {Teiga}}]{pogge12}
{Pogge}, R.~W., {Atwood}, B., {O'Brien}, T.~P., {et~al.} 2012, in Society of
  Photo-Optical Instrumentation Engineers (SPIE) Conference Series, Vol. 8446,
  Ground-based and Airborne Instrumentation for Astronomy IV, ed. I.~S.
  {McLean}, S.~K. {Ramsay}, \& H.~{Takami}, 84460G, \dodoi{10.1117/12.925693}

\bibitem[{{Prentice} {et~al.}(2016){Prentice}, {Mazzali}, {Pian}, {Gal-Yam},
  {Kulkarni}, {Rubin}, {Corsi}, {Fremling}, {Sollerman}, {Yaron}, {Arcavi},
  {Zheng}, {Kasliwal}, {Filippenko}, {Cenko}, {Cao}, \&
  {Nugent}}]{Prentice2016}
{Prentice}, S.~J., {Mazzali}, P.~A., {Pian}, E., {et~al.} 2016, \mnras, 458,
  2973, \dodoi{10.1093/mnras/stw299}

\bibitem[{{Rho} {et~al.}(2021){Rho}, {Evans}, {Geballe}, {Banerjee},
  {Hoeflich}, {Shahbandeh}, {Valenti}, {Yoon}, {Jin}, {Williamson}, {Modjaz},
  {Hiramatsu}, {Howell}, {Pellegrino}, {Vink{\'o}}, {Cartier}, {Burke},
  {McCully}, {An}, {Cha}, {Pritchard}, {Wang}, {Andrews}, {Galbany}, {Van Dyk},
  {Graham}, {Blinnikov}, {Joshi}, {P{\'a}l}, {Kriskovics}, {Ordasi}, {Szakats},
  {Vida}, {Chen}, {Li}, {Zhang}, \& {Yan}}]{rho2020bvc}
{Rho}, J., {Evans}, A., {Geballe}, T.~R., {et~al.} 2021, \apj, 908, 232,
  \dodoi{10.3847/1538-4357/abd850}

\bibitem[{{Sahu} {et~al.}(2018){Sahu}, {Anupama}, {Chakradhari}, {Srivastav},
  {Tanaka}, {Maeda}, \& {Nomoto}}]{Sahu2018}
{Sahu}, D.~K., {Anupama}, G.~C., {Chakradhari}, N.~K., {et~al.} 2018, \mnras,
  475, 2591, \dodoi{10.1093/mnras/stx3212}

\bibitem[{{Sauer} {et~al.}(2006){Sauer}, {Mazzali}, {Deng}, {Valenti},
  {Nomoto}, \& {Filippenko}}]{Sauer2006}
{Sauer}, D.~N., {Mazzali}, P.~A., {Deng}, J., {et~al.} 2006, \mnras, 369, 1939,
  \dodoi{10.1111/j.1365-2966.2006.10438.x}

\bibitem[{{Science Software Branch at STScI}(2012)}]{Pyraf}
{Science Software Branch at STScI}. 2012, {PyRAF: Python alternative for IRAF},
  Astrophysics Source Code Library, record ascl:1207.011.
\newblock \doeprint{1207.011}

\bibitem[{{Shahbandeh} {et~al.}(2022){Shahbandeh}, {Hsiao}, {Ashall}, {Teffs},
  {Hoeflich}, {Morrell}, {Phillips}, {Anderson}, {Baron}, {Burns}, {Contreras},
  {Davis}, {Diamond}, {Folatelli}, {Galbany}, {Gall}, {Hachinger}, {Holmbo},
  {Karamehmetoglu}, {Kasliwal}, {Kirshner}, {Krisciunas}, {Kumar}, {Lu},
  {Marion}, {Mazzali}, {Piro}, {Sand}, {Stritzinger}, {Suntzeff}, {Taddia}, \&
  {Uddin}}]{Shahbandeh_2022}
{Shahbandeh}, M., {Hsiao}, E.~Y., {Ashall}, C., {et~al.} 2022, \apj, 925, 175,
  \dodoi{10.3847/1538-4357/ac4030}

\bibitem[{{Shivvers} {et~al.}(2017){Shivvers}, {Modjaz}, {Zheng}, {Liu},
  {Filippenko}, {Silverman}, {Matheson}, {Pastorello}, {Graur}, {Foley},
  {Chornock}, {Smith}, {Leaman}, \& {Benetti}}]{Shivvers2017}
{Shivvers}, I., {Modjaz}, M., {Zheng}, W., {et~al.} 2017, \pasp, 129, 054201,
  \dodoi{10.1088/1538-3873/aa54a6}

\bibitem[{{Smartt}(2009)}]{Smartt2009}
{Smartt}, S.~J. 2009, \araa, 47, 63,
  \dodoi{10.1146/annurev-astro-082708-101737}

\bibitem[{{Smartt} {et~al.}(2015){Smartt}, {Valenti}, {Fraser}, {Inserra},
  {Young}, {Sullivan}, {Pastorello}, {Benetti}, {Gal-Yam}, {Knapic},
  {Molinaro}, {Smareglia}, {Smith}, {Taubenberger}, {Yaron}, {Anderson},
  {Ashall}, {Balland}, {Baltay}, {Barbarino}, {Bauer}, {Baumont}, {Bersier},
  {Blagorodnova}, {Bongard}, {Botticella}, {Bufano}, {Bulla}, {Cappellaro},
  {Campbell}, {Cellier-Holzem}, {Chen}, {Childress}, {Clocchiatti},
  {Contreras}, {Dall'Ora}, {Danziger}, {de Jaeger}, {De Cia}, {Della Valle},
  {Dennefeld}, {Elias-Rosa}, {Elman}, {Feindt}, {Fleury}, {Gall},
  {Gonzalez-Gaitan}, {Galbany}, {Morales Garoffolo}, {Greggio}, {Guillou},
  {Hachinger}, {Hadjiyska}, {Hage}, {Hillebrandt}, {Hodgkin}, {Hsiao}, {James},
  {Jerkstrand}, {Kangas}, {Kankare}, {Kotak}, {Kromer}, {Kuncarayakti},
  {Leloudas}, {Lundqvist}, {Lyman}, {Hook}, {Maguire}, {Manulis}, {Margheim},
  {Mattila}, {Maund}, {Mazzali}, {McCrum}, {McKinnon}, {Moreno-Raya},
  {Nicholl}, {Nugent}, {Pain}, {Pignata}, {Phillips}, {Polshaw}, {Pumo},
  {Rabinowitz}, {Reilly}, {Romero-Ca{\~n}izales}, {Scalzo}, {Schmidt},
  {Schulze}, {Sim}, {Sollerman}, {Taddia}, {Tartaglia}, {Terreran},
  {Tomasella}, {Turatto}, {Walker}, {Walton}, {Wyrzykowski}, {Yuan}, \&
  {Zampieri}}]{PESSTO}
{Smartt}, S.~J., {Valenti}, S., {Fraser}, M., {et~al.} 2015, \aap, 579, A40,
  \dodoi{10.1051/0004-6361/201425237}

\bibitem[{{Smith} {et~al.}(2006){Smith}, {Nordsieck}, {Burgh}, {Percival},
  {Williams}, {O'Donohue}, {O'Connor}, \& {Schier}}]{RSS}
{Smith}, M.~P., {Nordsieck}, K.~H., {Burgh}, E.~B., {et~al.} 2006, in
  \procspie, Vol. 6269, Society of Photo-Optical Instrumentation Engineers
  (SPIE) Conference Series, 62692A, \dodoi{10.1117/12.672415}

\bibitem[{{Soderberg} {et~al.}(2006){Soderberg}, {Nakar}, {Berger}, \&
  {Kulkarni}}]{Soderberg2006}
{Soderberg}, A.~M., {Nakar}, E., {Berger}, E., \& {Kulkarni}, S.~R. 2006, \apj,
  638, 930, \dodoi{10.1086/499121}

\bibitem[{{Soderberg} {et~al.}(2010){Soderberg}, {Chakraborti}, {Pignata},
  {Chevalier}, {Chandra}, {Ray}, {Wieringa}, {Copete}, {Chaplin},
  {Connaughton}, {Barthelmy}, {Bietenholz}, {Chugai}, {Stritzinger}, {Hamuy},
  {Fransson}, {Fox}, {Levesque}, {Grindlay}, {Challis}, {Foley}, {Kirshner},
  {Milne}, \& {Torres}}]{Soderberg2010}
{Soderberg}, A.~M., {Chakraborti}, S., {Pignata}, G., {et~al.} 2010, \nat, 463,
  513, \dodoi{10.1038/nature08714}

\bibitem[{{Stevance} {et~al.}(2017){Stevance}, {Maund}, {Baade}, {H{\"o}flich},
  {Howerton}, {Patat}, {Rose}, {Spyromilio}, {Wheeler}, \&
  {Wang}}]{Stevance2017}
{Stevance}, H.~F., {Maund}, J.~R., {Baade}, D., {et~al.} 2017, \mnras, 469,
  1897, \dodoi{10.1093/mnras/stx970}

\bibitem[{{Sukhbold} {et~al.}(2016){Sukhbold}, {Ertl}, {Woosley}, {Brown}, \&
  {Janka}}]{Sukhbold2016}
{Sukhbold}, T., {Ertl}, T., {Woosley}, S.~E., {Brown}, J.~M., \& {Janka}, H.~T.
  2016, \apj, 821, 38, \dodoi{10.3847/0004-637X/821/1/38}

\bibitem[{{Taddia} {et~al.}(2019){Taddia}, {Sollerman}, {Fremling},
  {Barbarino}, {Karamehmetoglu}, {Arcavi}, {Cenko}, {Filippenko}, {Gal-Yam},
  {Hiramatsu}, {Hosseinzadeh}, {Howell}, {Kulkarni}, {Laher}, {Lunnan},
  {Masci}, {Nugent}, {Nyholm}, {Perley}, {Quimby}, \& {Silverman}}]{Taddia2019}
{Taddia}, F., {Sollerman}, J., {Fremling}, C., {et~al.} 2019, \aap, 621, A71,
  \dodoi{10.1051/0004-6361/201834429}

\bibitem[{{Thomas} {et~al.}(2011){Thomas}, {Nugent}, \& {Meza}}]{Thomas2011}
{Thomas}, R.~C., {Nugent}, P.~E., \& {Meza}, J.~C. 2011, \pasp, 123, 237,
  \dodoi{10.1086/658673}

\bibitem[{{Tody}(1986)}]{IRAF1}
{Tody}, D. 1986, in Society of Photo-Optical Instrumentation Engineers (SPIE)
  Conference Series, Vol. 627, Instrumentation in astronomy VI, ed. D.~L.
  {Crawford}, 733, \dodoi{10.1117/12.968154}

\bibitem[{{Tody}(1993)}]{IRAF2}
{Tody}, D. 1993, in Astronomical Society of the Pacific Conference Series,
  Vol.~52, Astronomical Data Analysis Software and Systems II, ed. R.~J.
  {Hanisch}, R.~J.~V. {Brissenden}, \& J.~{Barnes}, 173

\bibitem[{{van Dokkum}(2001)}]{vanDokkum01}
{van Dokkum}, P.~G. 2001, \pasp, 113, 1420, \dodoi{10.1086/323894}

\bibitem[{{Vasylyev} {et~al.}(2022){Vasylyev}, {Filippenko}, {Vogl}, {Brink},
  {Brown}, {de Jaeger}, {Matheson}, {Modjaz}, {Mazzali}, {Patra}, {Smith}, {Van
  Dyk}, {Williamson}, {Yang}, {Zheng}, {deGraw}, {Fox}, {Gal-Yam}, {Jennings},
  \& {Rowe}}]{Vasylyev2022}
{Vasylyev}, S.~S., {Filippenko}, A.~V., {Vogl}, C., {et~al.} 2022, arXiv
  e-prints, arXiv:2203.08001.
\newblock \doarXiv{2203.08001}

\bibitem[{Virtanen {et~al.}(2020)Virtanen, Gommers, Oliphant, Haberland, Reddy,
  Cournapeau, Burovski, Peterson, Weckesser, Bright, {van der Walt}, Brett,
  Wilson, Millman, Mayorov, Nelson, Jones, Kern, Larson, Carey, Polat, Feng,
  Moore, {VanderPlas}, Laxalde, Perktold, Cimrman, Henriksen, Quintero, Harris,
  Archibald, Ribeiro, Pedregosa, {van Mulbregt}, \& {SciPy 1.0
  Contributors}}]{2020SciPy}
Virtanen, P., Gommers, R., Oliphant, T.~E., {et~al.} 2020, Nature Methods, 17,
  261, \dodoi{10.1038/s41592-019-0686-2}

\bibitem[{{Williamson} {et~al.}(2021){Williamson}, {Kerzendorf}, \&
  {Modjaz}}]{Williamson2021}
{Williamson}, M., {Kerzendorf}, W., \& {Modjaz}, M. 2021, \apj, 908, 150,
  \dodoi{10.3847/1538-4357/abd244}

\bibitem[{{Williamson} {et~al.}(2019){Williamson}, {Modjaz}, \&
  {Bianco}}]{williamson_2019}
{Williamson}, M., {Modjaz}, M., \& {Bianco}, F.~B. 2019, \apjl, 880, L22,
  \dodoi{10.3847/2041-8213/ab2edb}

\bibitem[{{Wong} {et~al.}(2006){Wong}, {Ryan-Weber}, {Garcia-Appadoo},
  {Webster}, {Staveley-Smith}, {Zwaan}, {Meyer}, {Barnes}, {Kilborn},
  {Bhathal}, {de Blok}, {Disney}, {Doyle}, {Drinkwater}, {Ekers}, {Freeman},
  {Gibson}, {Gurovich}, {Harnett}, {Henning}, {Jerjen}, {Kesteven}, {Knezek},
  {Koribalski}, {Mader}, {Marquarding}, {Minchin}, {O'Brien}, {Putman},
  {Ryder}, {Sadler}, {Stevens}, {Stewart}, {Stootman}, \& {Waugh}}]{Wong2006}
{Wong}, O.~I., {Ryan-Weber}, E.~V., {Garcia-Appadoo}, D.~A., {et~al.} 2006,
  \mnras, 371, 1855, \dodoi{10.1111/j.1365-2966.2006.10846.x}

\bibitem[{{Woosley} \& {Bloom}(2006)}]{wb2006}
{Woosley}, S.~E., \& {Bloom}, J.~S. 2006, \araa, 44, 507,
  \dodoi{10.1146/annurev.astro.43.072103.150558}

\bibitem[{{Yan} {et~al.}(2018){Yan}, {Perley}, {De Cia}, {Quimby}, {Lunnan},
  {Rubin}, \& {Brown}}]{Yan2018}
{Yan}, L., {Perley}, D.~A., {De Cia}, A., {et~al.} 2018, \apj, 858, 91,
  \dodoi{10.3847/1538-4357/aabad5}

\bibitem[{{Yaron} \& {Gal-Yam}(2012)}]{WISEREP}
{Yaron}, O., \& {Gal-Yam}, A. 2012, \pasp, 124, 668, \dodoi{10.1086/666656}

\bibitem[{{Yoon}(2015)}]{Yoon2015}
{Yoon}, S.-C. 2015, \pasa, 32, e015, \dodoi{10.1017/pasa.2015.16}

\bibitem[{{Yoon} \& {Langer}(2005)}]{Yoon2005}
{Yoon}, S.~C., \& {Langer}, N. 2005, \aap, 443, 643,
  \dodoi{10.1051/0004-6361:20054030}

\end{thebibliography}
\bibliographystyle{aasjournal}



\begin{deluxetable}{lcccccccc}
\tablecaption{Log of SN~2014ad Spectroscopic Observations \label{tab:spec}}
\tablehead{
\colhead{Observation Date} &
\colhead{Phase\tablenotemark{a}}  &
\colhead{Instrument} &
\colhead{Grating} &
\colhead{Wavelength} &
\colhead{Resolving Power} &
\colhead{Exp. Time} \\
\colhead{} &
\colhead{(rest-frame days)} &
\colhead{} &
\colhead{} &
\colhead{(\AA)} &
\colhead{($\lambda/\Delta\lambda$)} &
\colhead{(s)} &
\colhead{}
}
\startdata
2014 Mar 14 & $-4$ & SALT/RSS & G0900  & 3500$-$9410 & $1000$ & 1664 \\
2014 Mar 15 & $-3$ & SALT/RSS  & G0900 &3500$-$9410 & $1000$ & 2130 \\
2014 Mar 16 & $-2$ & SALT/RSS  & G0900 &3500$-$9410 & $1000$ & 2100 \\
2014 Mar 17 & $-1$ & SALT/RSS  & G0900 &3500$-$9410 & $1000$ & 2600 \\
2014 Mar 18 & $+0$ & SALT/RSS & G0900 & 3500$-$9410 & $1000$ & 1900 \\
2014 Mar 19 & $+1$ & HST/STIS/MAMA & G230L & 1615$-$3100 & 500 & 3478 \\
2014 Mar 19 & $+1$ & HST/STIS/CCD & G430L & 2900$-$5700 & 500 & 400 \\
2014 Mar 19 & $+1$ & HST/STIS/CCD & G750L & 5250$-$10230 & 500 & 400 \\
2014 Mar 19 & $+1$ & SALT/RSS & G0900 & 3500$-$9410 & $1000$ & 2448\\
2014 Mar 20 & $+2$ & SALT/RSS & G0900 & 3500$-$9410 & $1000$ & 2202\\
2014 Mar 21 & $+3$ & SALT/RSS & G0900 & 3500$-$9410 & $1000$ & 2352\\
2014 Mar 22 & $+4$ & SALT/RSS & G0900 & 3500$-$9410 & $1000$ & 2052\\
2014 Mar 25 & $+7$ & HST/STIS/MAMA & G230L & 1615$-$3100 & 500 & 4946 \\
2014 Mar 25 & $+7$ & HST/STIS/CCD & G430L & 2900$-$5700 & 500 & 700 \\
2014 Mar 25 & $+7$ & HST/STIS/CCD & G750L & 5250$-$10230 & 500 & 400 \\
2014 Mar 27 & $+9$ & SALT/RSS & G0900 & 3500$-$9410 & $1000$ & 1848\\
2014 Mar 28 & $+10$ & SALT/RSS & G0900 & 3500$-$9410 & $1000$ & 1948\\
2014 Mar 30 & $+12$ & SALT/RSS & G0900 & 3500$-$9410 & $1000$ & 1848 \\
2014 Apr 16\tablenotemark{b} & $+29$ & SALT/RSS & G0900 & 3500$-$9410 & $1000$ & 1649 \\
2014 Apr 18 & $+31$ & SALT/RSS & G0900 & 3500$-$9410 & $1000$ &  1664 \\
2014 Apr 20 & $+33$ & SALT/RSS & G0900 & 3500$-$9410 & $1000$ &  1388 \\
2014 May 01 & $+44$ & SALT/RSS & G0900 & 3500$-$9020 & $1000$ & 1500 \\
2014 May 12 & $+55$ & SALT/RSS & G0900 & 3500$-$9020 & $1000$ & 2200 \\
2014 May 17 & $+60$ & SALT/RSS & G0900 & 3500$-$9020 & $1000$ &  1500 \\
2014 May 23 & $+66$ & SALT/RSS & G0900 & 3500$-$9020 & $1000$ &  1500 \\
2014 Jun 04 & $+78$ & LBT/MODS & G400L/G670L & 3800$-$10000 & 3700/2875 & 3000\\
2014 Jun 12 & $+86$ & SALT/RSS & G0900 & 3500$-$9020 & $1000$ &  1500 \\
2014 Jun 27 & $+100$ & SALT/RSS & G0900 & 3500$-$9410 & $1000$ &  1500\\
2014 Jul 09 & $+112$ & SALT/RSS & G0900 & 3500$-$9410 & $1000$ & 1500\\
2014 Aug 02 & $+132$ & Magellan/IMACS & G600L & 4230$-$9400 & 1500 & 2400\\
2015 Jan 16 & $+302$ & Magellan/IMACS & G600L & 4230$-$9400 & 1500 & 5400\\
2015 Jan 20 & $+306$ & LBT/MODS & G400L/G670L & 3800$-$10000 & 3700/2875 & 3000\\
\enddata
\tablenotetext{a}{Phase is in rest-frame days relative to \textit{B}-maximum light (2014 Mar 18; MJD 56735.11).}
\tablenotetext{b}{Observations from Dartmouth College SALT program 2013-2-DC\_RSA-001 (PI: Fesen).}
\end{deluxetable}

\begin{deluxetable}{cccc}
\tablecaption{SN~2014ad Fe II 5169 \AA\ Velocities \label{tab:FeVels}}
    \tablehead{
    \colhead{Phase \textit{B}$_{\mathrm{max}}$} &
    \colhead{Phase \textit{V}$_{\mathrm{max}}$} &
    \colhead{Fe II Velocity} &
    \colhead{Reference}\\
    (rest-frame days) & (rest-frame days) & (10$^3$ \kms) & }
    \startdata
    +0 & -5.5 & 29.7  $\pm$ 4.55 & SALT (this work)\\
    +1 & -4.5 & 30.9  $\pm$ 4.82 & SALT (this work)\\
    +2 & -3.5 & 30.9  $\pm$ 4.82 & SALT (this work)\\
    +3 & -2.5 & 30.4  $\pm$ 4.70 & SALT (this work)\\
    +4 & -1.5 & 30.5  $\pm$ 4.70 & SALT (this work)\\
    +7 & +1.5 & 22.2  $\pm$ 2.82 & \citet{Sahu2018}\\
    +9 & +3.5 & 26.3  $\pm$ 2.19 & SALT (this work)\\
    +10 & +4.5 & 24.7  $\pm$ 2.19 & \citet{Sahu2018}\\
    +11 & +5.5 & 26.1  $\pm$ 2.07 & \citet{Sahu2018}\\
    +12 & +6.5 & 27.3  $\pm$ 2.04 & SALT (this work)\\
    +13 & +7.5 & 25.9  $\pm$ 1.18 & \citet{Sahu2018}\\
    +17 & +11.5 & 24.5  $\pm$ 2.70 & \citet{Sahu2018}\\
    +24 & +18.5 & 26.5  $\pm$ 1.44 & \citet{Stevance2017}\\
    +27 & +21.5 & 19.4  $\pm$ 1.49 & \citet{Sahu2018}\\
    +29 & +23.5 & 19.9  $^{+1.38}_{-1.49}$ & SALT (this work)\\
    +31 & +25.5 & 18.4  $^{+0.99}_{-0.88}$ & SALT (this work)\\
    +33 & +27.5 & 17.5  $\pm$ 0.79 & SALT (this work)\\
    +34 & +28.5 & 22.8  $^{+0.80}_{-0.86}$ & PESSTO \citep{PESSTO}\\
    +36 & +30.5 & 20.0  $\pm$ 1.14 & \citet{Sahu2018}\\
    +39 & +33.5 & 19.4  $\pm$ 1.38 & \citet{Stevance2017}\\
    +40 & +34.5 & 18.7  $\pm$ 1.40 & \citet{Sahu2018}\\
    +43 & +37.5 & 20.0  $\pm$ 2.05 & PESSTO \citep{PESSTO}\\
    +44 & +38.5 & 14.2  $\pm$ 2.04 & SALT (this work)\\
    +46 & +40.5 & 16.4  $\pm$ 1.68 & \citet{Sahu2018}\\
    +49 & +43.5 & 20.0  $\pm$ 0.09 & \citet{Stevance2017}\\
    +55 & +49.5 & 15.3  $\pm$ 0.74 & SALT (this work)\\
    +60 & +54.5 & 11.3  $\pm$ 0.82 & SALT (this work)\\
    +64 & +58.5 & 16.6  $\pm$ 1.09 & \citet{Stevance2017}\\
    +66 & +60.5 & 12.7  $\pm$ 0.88 & SALT (this work)\\
    +72 & +66.5 & 17.2  $\pm$ 1.28 & \citet{Stevance2017}\\
    +78 & +72.5 & 11.9  $\pm$ 1.03 & \citet{Sahu2018}\\
\enddata
\end{deluxetable}

\end{document}